\documentclass[fleqn,usenatbib]{mnras}
\pdfoutput=1
\usepackage{newtxtext,newtxmath}

\usepackage[T1]{fontenc}
\usepackage{ae,aecompl}

\usepackage{graphicx}	
\usepackage{amsmath}	
\usepackage{amssymb}	

\usepackage{caption}
\usepackage{subfig}
\def\code#1{\texttt{#1}}

\title[AGN-driven quenching of satellite galaxies]{AGN-driven quenching of satellite galaxies}

\author[G. Dashyan et al.]{
Gohar Dashyan$^{1}$\thanks{E-mail: dashyan@iap.fr},
Ena Choi$^{2}$,
Rachel S. Somerville$^{3,4}$,
Thorsten Naab$^{5}$,
\newauthor
Amanda C. N. Quirk$^{6}$,
Michaela Hirschmann$^{1,7}$
and Jeremiah P. Ostriker$^{2,8}$
\\
$^{1}$Sorbonne Universit\'e, UPMC-CNRS, UMR7095, Institut d'Astrophysique de Paris, F-75014 Paris, France\\
$^{2}$Department of Astronomy, Columbia University, 550 W 120th Street, New York, NY 10027, USA\\
$^{3}$Department of Physics and Astronomy, Rutgers, The State University of New Jersey\\
$^{4}$Center for Computational Astrophysics, Flatiron Institute, 162 5th Avenue, New York, NY, 10010, USA\\
$^{5}$Max-Planck-Institut f\"ur Astrophysik, Karl-Schwarzschild-Strasse 1, 85741 Garching, Germany\\
$^{6}$UCO/Lick Observatory, University of California at Santa Cruz
1156 High Street Santa Cruz, CA 95064, USA\\
$^{7}$University of Vienna, Institute for Astronomy, T\"urkenschanzstrasse 17, 1180 Vienna, Austria\\
$^{8}$Department of Astrophysical Sciences, Princeton University, Princeton, NJ 08544, USA
}

\date{Accepted XXX. Received YYY; in original form ZZZ}

\pubyear{2019}

\begin{document}
\label{firstpage}
\pagerange{\pageref{firstpage}--\pageref{lastpage}}
\maketitle

\begin{abstract}
We explore the effect of active galactic nucleus (AGN) feedback from central galaxies on their satellites by comparing two sets of cosmological zoom-in runs of 27 halos with masses ranging from $10^{12}$ to $10^{13.4}$ $\rm{M}_{\odot}$ at $z=0$, with (wAGN) and without (noAGN) AGN feedback. Both simulations include stellar feedback from multiple processes, including powerful winds from supernovae, stellar winds from young massive stars, AGB stars, radiative heating within Str\"omgren spheres and photoelectric heating. Our wAGN model is identical to the noAGN model except that it also includes a model for black hole seeding and accretion, as well as AGN feedback via high-velocity broad absorption line winds and Compton/photoionization heating. We show that the inclusion of AGN feedback from the central galaxy significantly affects the star formation history and the gas content of the satellite galaxies. AGN feedback starts to affect the gas content and the star formation of the satellites as early as $z=2$. The mean gas rich fraction of satellites at $z=0$ decreases from 15\% in the noAGN simulation to 5\% in the wAGN simulation. The difference between the two sets extends as far out as five times the virial radius of the central galaxy at $z=1$. We investigate the quenching mechanism by studying the physical conditions in the surroundings of pairs of satellites matched across the wAGN and noAGN simulations and find an increase in the temperature and relative velocity of the intergalactic gas.
\end{abstract}




\section{Introduction}

The process by which star formation ceases in galaxies is called \textit{galaxy quenching}. The quenching mechanisms, i.e. the processes that prevent gas from cooling and/or forming stars, as well as their relative importance, are still a topic of active investigation.

Theoretical models of galaxy formation implemented in semi-analytical models and cosmological hydrodynamical simulations are now able to reproduce fairly well the observed stellar mass functions and luminosity functions of galaxies (see \citealt{Somerville15,NaabOstriker17} for recent reviews). This good agreement has been obtained by tuning free parameters and the use of subgrid recipes for baryonic physics such as star formation, stellar feedback and active galactic nucleus (AGN) feedback. Another more stringent and challenging constraint for galaxy formation and evolution models is given by the statistical correlations of quiescent galaxies. \cite{Peng10} suggested that two distinct processes are operating, ``mass quenching'', which is independent of environment, and  ``environment quenching'', which is independent of internal properties such as stellar mass. \cite{Peng12} suggested that the fraction of quiescent centrals depends only on stellar mass, whereas the fraction of quiescent satellites depends on both mass and environment. \cite{Woo2013} argue that the quenched fraction of galaxies at
$z=0-1$ seems to be primarily driven by the dark matter halo mass. They show that the passive fraction of central galaxies is more correlated with halo mass at fixed stellar mass than with stellar mass at fixed halo mass. For satellite galaxies, there is a strong dependence on both halo mass and distance to the halo center. 
 
Various environmental processes, gravitational or hydrodynamical, could quench satellite galaxies. Tidal forces from the host halo can strip mass from the satellite \citep{Dekel03,Wetzel10}, and frequent high speed encounters with neighboring galaxies can tidally heat satellites \citep{Farouki81,Moore98}. Ram pressure from the hot gas in the host halo and the high orbital velocity of the satellite can also strip or heat the gas from the satellite \citep{Gunn72,Tonessen09}.  Strangulation or starvation, a more gradual process, is the lack of accretion of new gas \citep{Larson80}. In semi-analytical models, environmental processes are often described with simplified recipes where galaxies stop accreting new gas from the hot halo or the intergalactic medium once they become satellites. This prescription overproduces the fraction of quiescent satellites \citep{Kimm09}. \cite{Hirschmann14} showed that even with a delayed strangulation model, the quiescent satellite fractions are significantly over-estimated. The relative importance of these environmental processes is still unclear. So is their significance with respect to internal processes, since these satellites are also affected by feedback from stars and potential AGN, which can even enhance the environmental processes \citep{Bahe15}. Furthermore, in the hierarchical structure formation scenario, galaxies join more and more massive systems, experiencing various environmental histories during their lifetime so that external and internal processes are connected and hard to disentangle. 

Several studies have claimed to detect a correlation between properties of galaxies, such as morphology, gas content, star formation rate, and those of neighboring galaxies. This effect, called \textit{galactic conformity} was first observed for satellites of larger central galaxies: satellites of passive host galaxies are more likely to also be passive relative to their counterparts around star-forming hosts, at fixed group mass \citep{Weinmann06,Kauffmann10,Wang12, Phillips14,Knobel15}. This suggests that quenching mechanisms for central galaxies also impact the satellite galaxies. A detection of a conformity signal on projected distances of up to 4 Mpc was presented in \cite{Kauffmann13}. However, it should be noted that subsequent studies have questioned the validity of past detections due to selection biases or possible errors in estimating halo masses (\citealt{Campbell15}; \citealt{Sin17}; \citealt{Calderon18}; \citealt{Tinker18} ; see section 6.2.2 of \citealt{Wechsler18} for a recent review). The physical origin of galactic conformity is still unclear. \cite{Hearin15,Hearin16} suggest that the large-scale conformity signal is a detection of central galaxy assembly bias, \textit{i.e.} the fact that the formation histories of dark matter halo are spatially correlated. Another physical explanation, suggested by \cite{Kauffmann15,Kauffmann18}, is that gas is heated over large scales at early times by AGN feedback. Recently \cite{Shen2019} observed a possible evidence of AGN-driven quenching of neighboring galaxies at $z \sim 1$, using data from the ORELSE survey \citep{Lubin2009}. \cite{Uchiyama2019} found that Lyman alpha emitters with high rest-frame equivalent width of Ly$\alpha$ emission with low stellar mass ($< 10^8 \rm{M}_{\odot} $), are predominantly scarce in the quasar proximity zones.

In the context of the formation of massive elliptical galaxies, observations have established that early-type galaxies form and become red and dead early but continue to grow in mass and size without much late in-situ star formation  \citep[e.g.][]{Daddi05,Trujillo06,Buitrago06,Szomoru09,Porter14,Choi18}. These observations favor a two-phase formation scenario \citep{Naab07,Oser10}. In this scenario, the progenitors build the bulk of their mass in short but intense starbursts events at $z>2$. Then, a progressive process of mergers  with satellites and accretion of old stars produces the stellar envelopes resulting in a significant size growth \citep{Bezanson09,vanDokkum10,Hirschmann13,Sales12}. Therefore, the gas and stellar content of these merging satellites is an essential parameter for the size growth and the amount of gas available for late star formation in this two-phase formation scenario. 

Recent hydrodynamical simulations by \cite{Choi17,Choi18} showed that the inclusion of AGN feedback effectively quenches the star formation in massive galaxies, transforming blue compact galaxies into quiescent galaxies. AGN feedback also removes and prevents new accretion of cold gas, shutting down in-situ star formation and causing subsequent mergers to be gas-poor. Gas poor minor mergers then build up an extended stellar envelope. AGN feedback also puffs up the central region of galaxies through the fast AGN driven winds as well as the slow expulsion of gas while the black hole (BH) is quiescent. In the simulation run without AGN feedback, large amounts of gas accumulate in the central region of galaxies, triggering star formation and leading to overly massive blue galaxies with dense stellar cores. As shown by \cite{Hirschmann17}, this strongly reduced star formation rate turns out to be necessary in order to be consistent with the observed offset in $[O_{\rm{III}}]\rm{\lambda} 5007/\rm{H\beta}$ for a given $[N_{\rm{II}}]\rm{\lambda} 6584/\rm{H\alpha}$ in high-redshift galaxies.

Do the massive black holes in the central galaxies affect the satellite systems? To our knowledge, this has never before been explicitly investigated in numerical cosmological simulations. In this paper, we use the two sets of cosmological zoom-in simulations as described \cite{Choi17} to explore the effect of AGN feedback from central galaxies on their satellites or neighboring galaxies. We compare two zoom-in runs of of 27 halos with virial masses that range from $10^{12}$ to $10^{13.4}$ $M_{\odot}$, with (wAGN) and without (noAGN) AGN feedback. We show how the star formation in these satellite galaxies is efficiently quenched by the central AGN.

This paper is structured as follows. In Section \ref{simulations and methods}, we describe the simulations used as well as our halo finding and tracking methods, and in Section \ref{results}, we present the results of our analysis. In Section \ref{physics}, we investigate the physical mechanisms at the origin of the additional quenching of the satellites observed in the wAGN simulation. We summarize and discuss our results in Section \ref{Conclusion}.

\section{Simulations and methods}
\label{simulations and methods}

In this section, we give a brief overview of the physics relevant to our study. A more detailed description of the simulations can be found in \cite{Choi17}. 

\subsection{Code basics and setup}

The simulations are run with a modified version of the parallel smoothed particle hydrodynamics (SPH) code GADGET-3 \citep{Springel05}, SPHGal \citep{Hu14}, that includes a density-independent pressure-entropy SPH formulation \citep{Hopkins13}. To further improve over standard SPH, we adopt the Wendland $C^4$ kernel with 200 neighboring particles. We also include the improved artificial viscosity implementation presented by \cite{Cullen10} and an artificial thermal conductivity according to \cite{Read12} in order to reduce the noise in pressure estimates in the presence of strong shocks. Finally, a time-step limiter is employed according to \cite{Saitoh09} and \cite{Durier12} to ensure that neighboring particles have similar time steps and that ambient particles do not remain inactive when a shock is approaching.

\subsection{Star formation and stellar/supernova feedback}

Star formation and chemical evolution are modeled as described in \cite{Aumer13}, which allows chemical enrichment by winds driven by Type I SNe, Type II SNe, and asymptotic giant branch (AGB) stars. Eleven species of metals are tracked explicitly, and the net cooling rates are calculated based on temperature, density of gas and individual element abundances. We adopted the cooling rate from \cite{Wiersma09} for optically thin gas in ionisation equilibrium. Redshift-dependent UV/X-ray and cosmic microwave backgrounds with a modified \cite{Haardt12} spectrum are also included.

Stars are formed stochastically if the gas density exceeds a density threshold. This threshold is given as $n_{\rm th} \equiv n_{0} \left(T_{\rm gas} / T_{0}\right)^3 \left(M_{0}\right / M_{\rm gas})^2$ with $n_0 = 2.0 \rm{cm}^{-3}$ and $T_0 = 12,000 \rm{K}$, and $M_0$ is the gas particle mass. The star formation rate is calculated as $\rm{d} \rho_*/\rm{d}t = \eta \rho_{\rm gas}/t_{\rm dyn}$, where $\rho_*$, $\rho_{\rm gas}$, and $t_{\rm dyn}$ are the stellar density, gas density, and local dynamical time for the gas particle, respectively. The star formation efficiency $\eta$ is set to 0.025.

Stellar feedback is included in the form of stellar winds and heating by ionizing radiation from young massive stars. Momentum from stellar winds is added to the surrounding gas particles, while cold gas within the Str\"omgren radius of hot stars is heated to $T = 10^4 \,\rm{K}$.

In our supernova (SN) feedback model, a single SN event is assumed to eject mass in an outflow with a velocity $v_{\rm out,SN} = 4500 \, \rm{km\,  s}^{-1}$, a typical velocity of outflowing materials in SNe. SN energy and momentum are distributed to the surrounding interstellar medium (ISM) from the SN event. Depending on the distance to the SN events, each nearby gas particle is affected by one of three successive phases of SN remnant (SNR) evolution: the ejecta-dominated free expansion (FE) phase, the energy-conserving Sedov-Taylor blast-wave SNR phase, and the momentum-conserving snowplow phase. SN energy is transferred by conserving the ejecta momentum for gas particles within the radius of the FE phase. For gas particles lying outside the FE radius but within the Sedov-Taylor phase, the SN energy is transferred as 30\% kinetic and 70\% thermal. Finally, at larger radii in the snowplow phase, only a fraction of the original SN energy is transferred as radiative cooling becomes significant. See Appendix A of \cite{Nunez17} for a detailed description  of the implementation of the SN feedback model.

Feedback from low and intermediate initial mass stars via slow winds during an AGB phase is also included. Momentum and energy from old star particles are transferred to the neighboring gas particles in a momentum-conserving way. The outflowing wind velocity of AGB stars is assumed to be $v_{\rm out,AGB} = 10 \,\rm{km\,  s}^{-1}$, corresponding to typical outflowing velocities of AGB stars \citep{Nyman92}. Metal-enriched gas from all of these prescriptions is continuously added to the ISM. Metal diffusion, which allows for the mixing and spreading of metals in the enriched gas, is also included.

\subsection{Black hole growth and feedback}
In the simulations, the BHs are treated as collisionless sink particles and are seeded in newly forming dark matter halos. Dark matter halos are identified on the fly during a simulation by a friends-of-friends algorithm. The new BHs are seeded with mass of $10^{5.15}   \rm{M}_{\odot}$ such that any halo above $10^{11.15} \rm{M}_{\odot}$ contains one BH at its center if it does not already have a BH. The BH mass can then grow by gas accretion or by merging, when the two BHs fall within each other's local SPH smoothing lengths and their relative velocities are smaller than the local sound speed. Gas accretion onto the BH follows a Bondi-Hoyle-Littleton parametrization \citep{Bondi52}. 

The soft Bondi criterion introduced in \cite{Choi12} is also included to avoid the unphysical accretion of unbound gas from outside the Bondi radius of the BH. This criterion statistically limits the accretion to the gas within the Bondi radius. It also accounts for the size of the gas particle as the physical properties of each gas particle are smoothed within the kernel size in the smoothed particle hydrodynamics simulations. Full accretion is only allowed when the total volume of a gas particle is included within the Bondi radius. If a gas particle volume is partially included within the Bondi radius, its probability of being absorbed by the BH is reduced. Finally, in order to account for the time that it takes to a particle to be accreted onto a BH, the free-fall timescale is included following \cite{Choi12}.

Our AGN feedback model \citep{Ostriker2010,Choi12,Choi14} consists of two main components: 
\begin{enumerate}
\item Mechanical feedback as in the broad absorption line winds, which carry energy, mass, and momentum into the surrounding gas. Winds are launched
from the central region around the BH, with a fixed wind velocity of $v_{\rm outf,AGN} = 10,000 \,\rm{km\,  s}^{-1}$. The total energy flux carried by the wind is $\dot{E}_{\rm w} = \epsilon_{\rm w} \dot{M}_{\rm acc}c^2$, where the efficiency parameter $\epsilon_{\rm w}$ is set to 0.005 \citep{Choi17}, $\dot{M}_{\rm acc}$ is the mass accretion rate onto the BH and $c$ is the speed of light. The mass flux and momentum flux carried by the wind are $\dot{M}_{\rm outf} =2\epsilon_{\rm w} \dot{M}_{\rm acc} c^2/v_{\rm outf,AGN}^2 $ and $\dot{p}_{\rm outf} = 2\epsilon_{\rm w}\dot{M}_{\rm acc}  c^2/v_{\rm outf,AGN} $, respectively. Thus, for our selected feedback efficiency $\epsilon_{\rm w}$ and wind velocity $v_{\rm outf,AGN}$, we have $\dot{M}_{\rm outf} =9\dot{M}_{\rm acc}$. Therefore 90\% of mass entering the central region is expelled, and 10\%  the inflowing mass is accreted onto the BH. The wind particles are stochastically selected among the particles entering the central region.
The selected gas particles receive the wind kick in a direction parallel or antiparallel to their angular momentum vectors. The emitted wind particle shares its momentum with its two nearest neighbors to reproduce the shock-heated momentum-driven flows. The residual energy increases the temperature of the impacted gas particles; therefore, the total energy and momentum are conserved. This prescription gives a ratio of kinetic to thermal energy in the outflowing particles similar to that in the standard Sedov-Taylor blast wave.

\item Radiative feedback via the Compton and photoionization heating from the X-ray radiation from the accreting BH, the radiation pressure associated with the heating, and the Eddington force. The emergent AGN spectrum and metal-line heating are taken from \cite{Sazonov04}. X-ray radiation is coupled to
the surrounding gas according to \cite{Sazonov05}. The radiation pressure on each gas element is also calculated. Accretion onto the BHs is not limited; the Eddington force is included, acting on electrons in the neighboring gas through the hydrodynamic equations, directed radially away from BH. In this way we allow that super-Eddington gas accretion occasionally occurs, so that the corresponding feedback effect naturally reduces the inflow and increases the outflow. Also included are metallicity-dependent heating prescriptions due to photoelectric emission and metal line absorption. 
\end{enumerate}

We refer the reader to \cite{Choi17} for more details about our feedback prescriptions.
See also \cite{Brennan18} for the analysis of the wind properties and gas cycle in these simulations; \cite{Frigo18} for the analysis of stellar kinematics; \cite{Hirschmann17,Hirschmann18} for the analysis of optical and UV emission line properties.

\subsection{Zoom simulations}

The initial conditions for the zoom-in simulations are described in \cite{Oser10,Oser12}. The halos of interest are selected from a dark matter only simulation using a flat cosmology with parameters obtained from from WMAP3 (\citealt{Spergel}; $h=0.72$, $\Omega_{\rm b}=0.044$, $\Omega_{\rm dm}=0.216$, $\Omega_{\Lambda}=0.74$, $\sigma_{8}=0.77$, $n_s = 0.95$). At any given snapshot we trace back all particles close to the halos of interest from redshift zero. We replace those particles with higher-resolution gas and dark matter particles. Then, new high-resolution initial conditions are simulated from redshift z = 43 to z = 0.

The simulations have been performed at two resolutions. The reference resolution has a mass resolution for the star and
gas particles of $m_{*,\rm{gas}}=5.8 \times 10^6 \, {\rm M}_{\odot}$ , and dark matter particles have $m_{\rm{dm}}=3.4 \,\times\, 10^7 \, {\rm M}_{\odot}$. We use the comoving gravitational softening lengths $\varepsilon_{\rm{gas,star}} = 400 \,\rm{pc}\, \rm{h}^{-1}$ for the gas and star particles and  $\varepsilon_{\rm halo} = 890 \,\rm{pc}\, \rm{h}^{-1}$  for dark matter. The high-resolution simulations have been performed with eight times better mass resolution than the reference resolution, with $m_{*,\rm{gas}}=7.4\, \times 10^5 \, {\rm M}_{\odot}$ and $m_{\rm{dm}}=4.3 \times 10^6 \, {\rm M}_{\odot}$, and twice better spatial resolution with $\epsilon_{\rm gas,star} = 200 \, \rm{pc}\, \rm{h}^{-1}$ and  $\epsilon_{\rm halo} = 450 \,\rm{pc}\, \rm{h}^{-1}$. In the \hyperref[appendix]{Appendix}, we discuss the resolution convergence.

The simulated halo masses are in the range $1.4 \times 10^{12} \rm{M}_{\odot} \leq M_{\rm vir} \leq  2.3 \times 10^{13} \rm{M}_{\odot}$  at z = 0, and the stellar masses of central galaxies are $8.2 \times 10^{10} \rm{M}_{\odot}  \leq  M_{*}  \leq  1.5 \times 10^{12} \, \rm{M}_{\odot}$ at present day. The virial radii at $z=0$ are in the range $ 300\, \rm{kpc} \leq R_{\rm vir} \leq  700 \, \rm{kpc}$. 

Dark matter halos were found using the ROCKSTAR algorithm from \cite{Behroozi13}. Halo masses were calculated using
a spherical overdensity threshold fixed at 200 times that of the critical density at the considered redshift. The merger trees were computed using the CONSISTENT TREES \citep{consistenttrees}.

In what follows, we will examine the satellite galaxies in zoom regions run with the two different models, wAGN and noAGN.

\section{Results}
\label{results}
Here we present the results of our analysis. First, in section \ref{fixed z} we examine the properties of the satellite galaxies at various redshifts. Then, in section \ref{histories} we follow back in time the main progenitors of the satellites found at redshift 0. Since we want to understand the effect of the central AGN, we exclude from our study the satellites in the wAGN simulations in which a BH has been placed, which could be quenched compared to their counterparts in the noAGN model independently from the central AGN. As a consequence, we also exclude from the noAGN sample the satellites that are above the mass threshold for BH seeding, in order to use the noAGN model as a control sample and to isolate the effect of the central AGN. We use a lower mass cut of 64 dark matter particles for selecting the satellites. These lower (given by the number of dark matter particles) and higher (given by the mass limit for BH seeding) mass cuts give satellite masses in the range $2 \times 10^9 \rm{M}_{\odot} \leq M_{\rm vir, satellites} \leq  10^{11.15} \rm{M}_{\odot} $. We refer to the main halo of each zoom in simulation as the central galaxy even though we sometimes extend our study to neighboring galaxies that lie outside the virial radius of the central galaxy. We use the \code{pygad} tool \citep{pygad} for our analysis.

\subsection{Properties of the satellite galaxy population at various redshifts}
\label{fixed z}

\subsubsection{Gas content and star formation}

\begin{figure}
    \centering
    \includegraphics[width=\columnwidth]{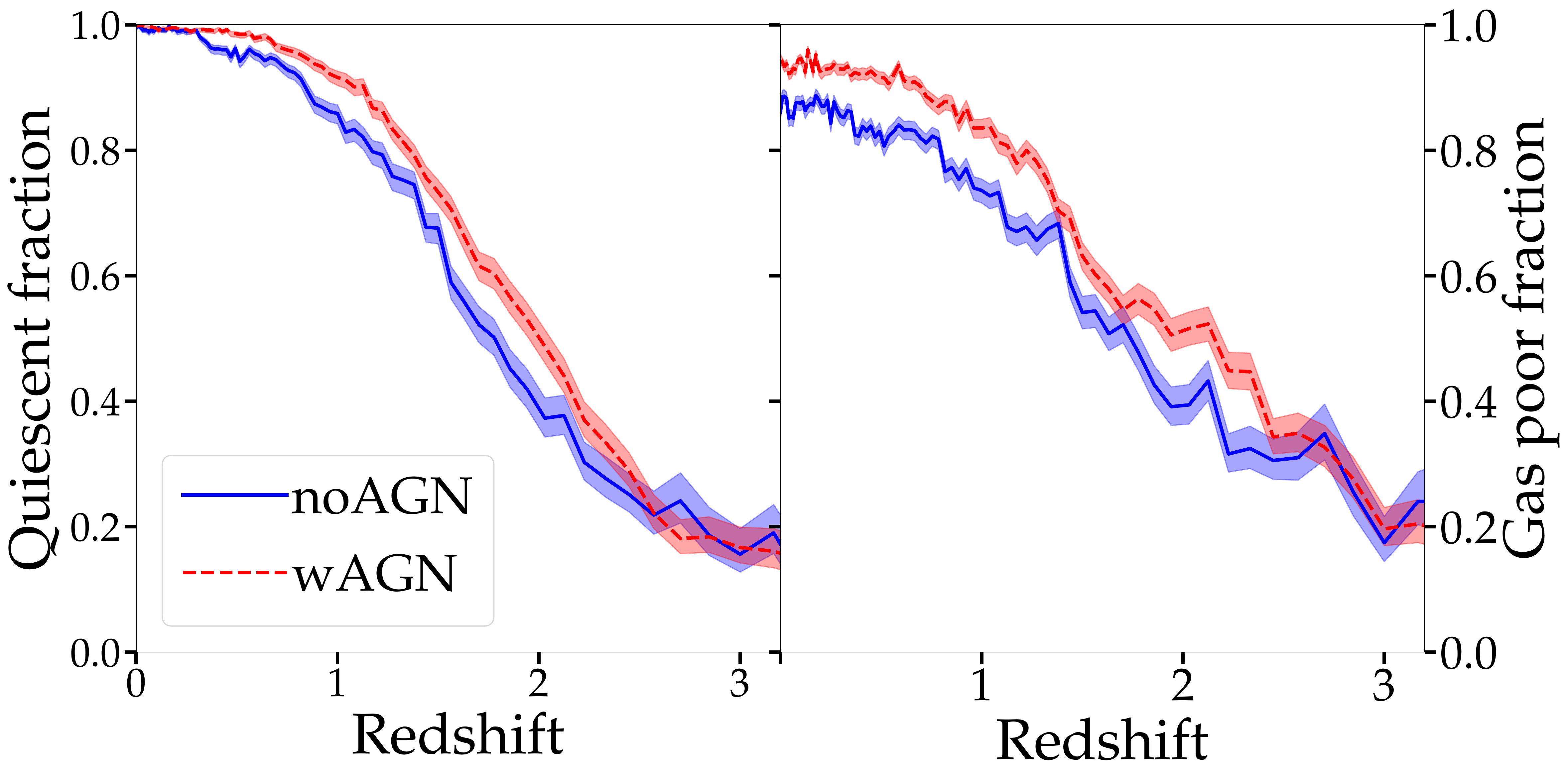}
    \caption{Comparison of the quiescent and gas poor satellite fractions in the wAGN and noAGN simulations. Each redshift bin contains a few hundreds of satellites ($\sim 1000$ at $z=0$ and $\sim 100$ at $z=3$). \textbf{Left}: Quiescent fraction (sSFR $< 0.3/t_{\rm H}$, where $t_{\rm H}$ is the age of the universe at each redshift) of the satellite galaxies within the virial radius of the central as a function of redshift. \textbf{Right}: Gas poor fraction ($M_{\rm gas}/(M_{\rm stars} + M_{\rm{gas}}) < 0.1 $) of the satellite galaxies within the virial radius of the central as a function of redshift. The error bars indicate the $68$\% confidence intervals computed for a beta distribution.}
 \label{quiescentgas_vs_z}
\end{figure}

\begin{figure}
    \centering
    \includegraphics[width=\columnwidth]{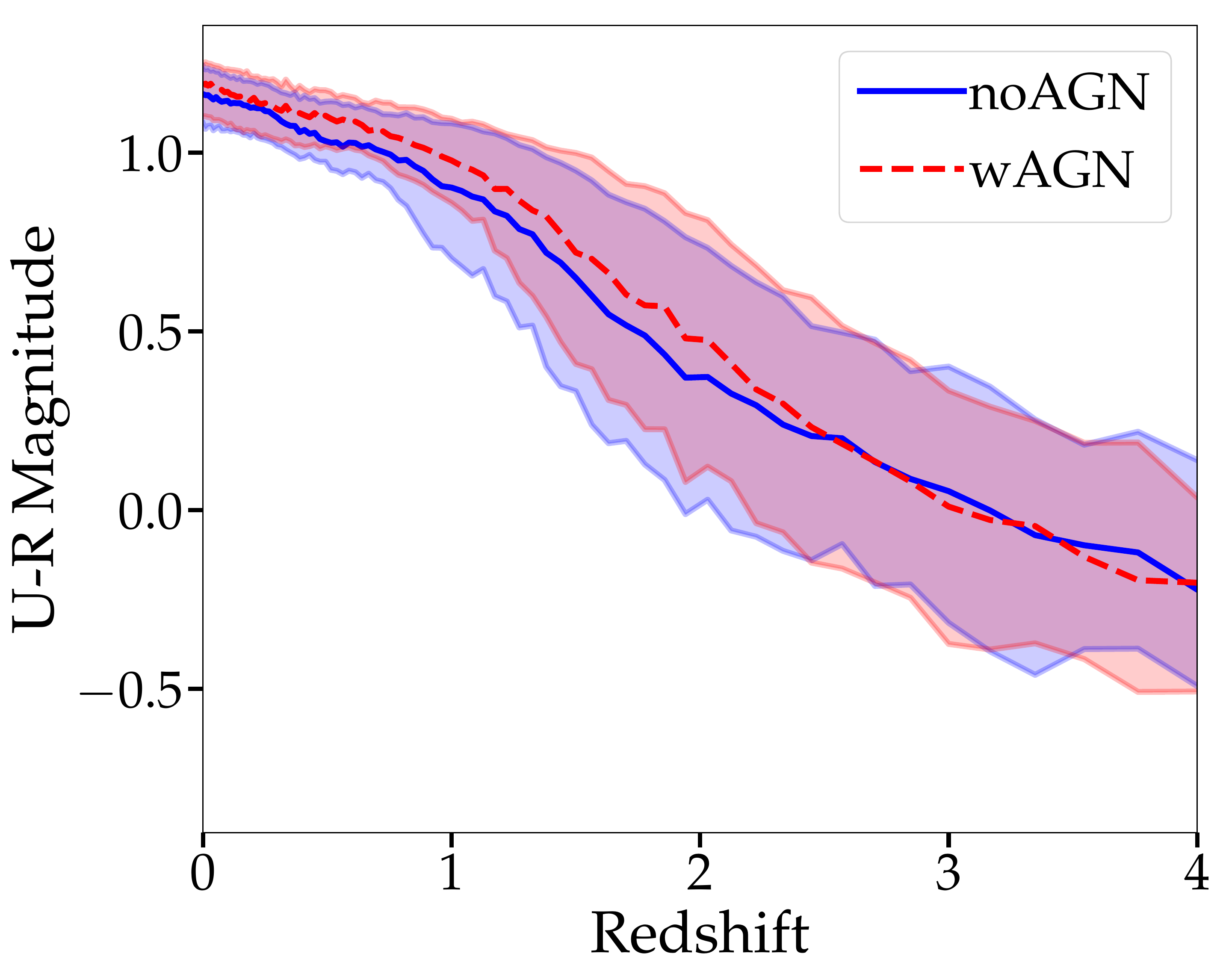}
    \caption{Comparison of the U - R magnitude of the satellite galaxies in the wAGN and noAGN simulations. We stacked the satellites of the 27 zoom in simulations. Each redshift bin contains a few hundreds of satellites ($\sim 1000$ at $z=0$ and $\sim 100$ at $z=3$). The shaded area shows the 16\% and the 84\% percentiles of the distributions.}
 \label{colors}
\end{figure}

\begin{figure*}
    \centering
  \includegraphics[width=\textwidth]{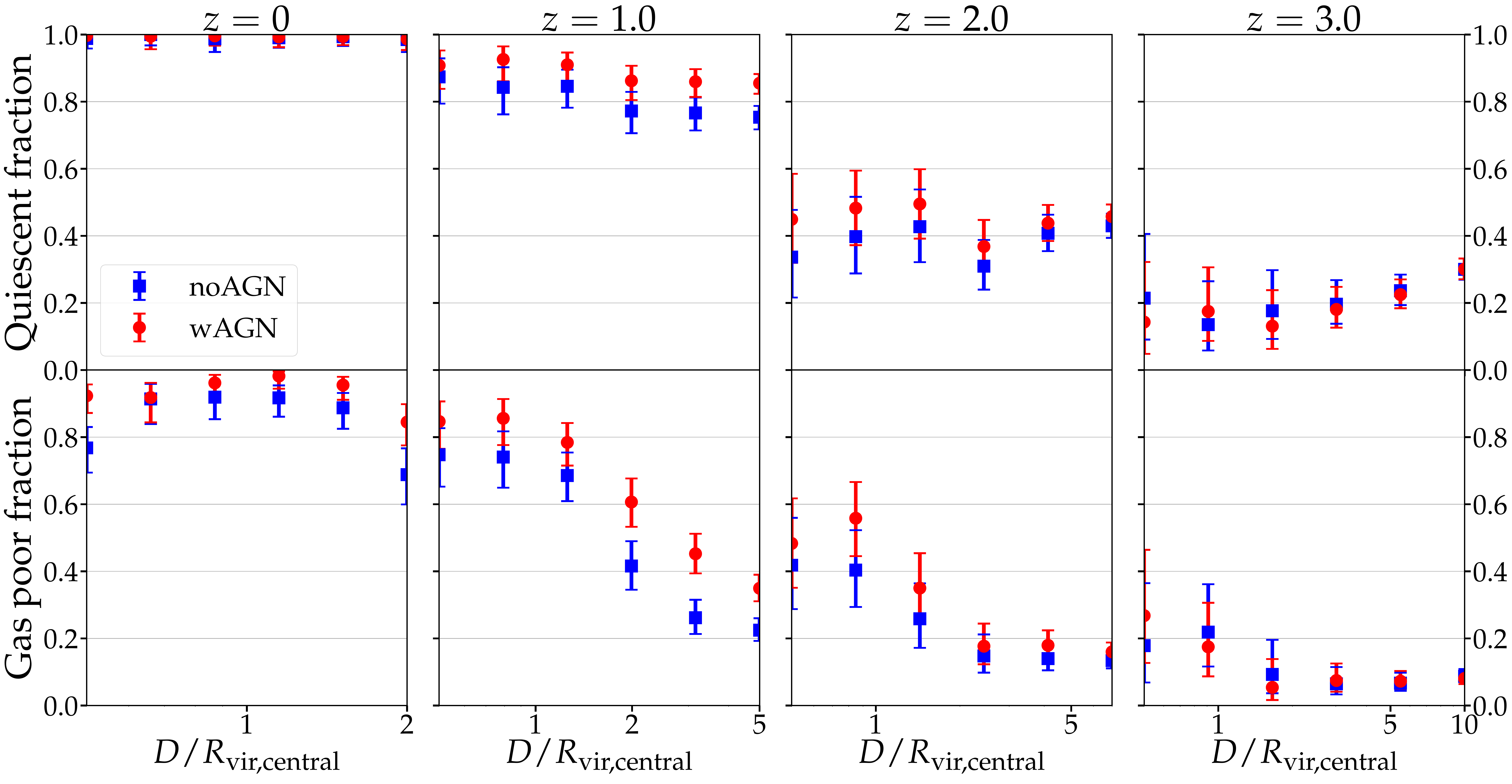}
    \caption{Quiescent and gas poor satellite fractions of the galaxies around the central as a function of the normalized distance to the central (distance to the central divided by the virial radius of the central). \textbf{Top} : Quiescent fractions (sSFR $< 0.3/t_{\rm H}$, where $t_{\rm H}$ is the age of the universe at each redshift), the different panels show different redshifts. \textbf{Bottom}: Gas poor fractions ($M_{\rm gas}/(M_{\rm stars} + M_{\rm gas}) < 0.1 $). We stacked the 27 zoom in simulations. The wAGN and noAGN samples each contain $\sim 1700$ at $z=0$ and $3000$ at $z=0,1,2$ (the galaxies are taken within a sphere of $2,5,7$ and $10$ times the virial radius of the central at $z=0,1,2,3$ respectively). The error bars indicate the $99.7$\% confidence intervals computed for a beta distribution. The quiescent and gas poor fractions are similar at $z=3$ and the difference grows at lower redshifts, first out to twice the virial radius of the central galaxy at $z=2$, then out to more than five times the virial radius of the central galaxy at $z=1$.}
  \label{quiescentgas_vs_distance_various_z}
\end{figure*}

In this section, we study the evolution of the satellite population around the central galaxies across cosmic time. Fig.~\ref{quiescentgas_vs_z} shows the overall quiescent and gas poor fractions of the satellite galaxies as a function of redshift: for each given redshift, we identified in each zoom-in simulation the satellite galaxies within one virial radius of the main halo center and computed the gas poor and quiescent fractions. Each redshift bin ($\sim 100$ Myr) contains a few hundreds of satellites ($\sim 1000$ at $z=0$ and $\sim 100$ at $z=3$, in both wAGN and noAGN simulations). We divide our galaxies into star forming and quiescent galaxies based on the commonly adopted dividing line at sSFR $< 0.3/t_{\rm H}$, where $t_{\rm H}$ is the age of the universe at each redshift (e.g. \citealt{Franx08}). We choose a gas fraction threshold of $0.1$ for being gas poor, the gas fraction being the ratio of the mass in gas to the mass in gas and stars ($M_{\rm gas}/(M_{\rm stars} + M_{\rm gas}) < 0.1 $), where $M_{\rm gas}$ and $M_{\rm stars}$ are measured within one tenth of the virial radius of the satellites. The error bars indicate the $68$\% confidence intervals computed as given by a beta distribution \citep{Cameron11}. Starting from $z=2$ the difference between the wAGN and noAGN simulations grows. The quiescent fractions seem to be most significantly different from $z=2$ to $z=1$. At $z=0$, the wAGN and noAGN satellites have quiescent fractions close to one. The gas poor fractions are different from $z=2$ and the difference between the noAGN and the wAGN keeps growing until $z=0$. The mean gas rich fraction at $z=0$ decreases from 15\% in the noAGN simulation to 5\% in the wAGN simulation.

Fig~\ref{colors} shows the $U - R$ magnitude of the satellites computed using a \code{pygad} module that reads and interpolates \cite{Bruzual03} single stellar population model. The effects of dust attenuation are not included. The shaded area shows the 16\% and the 84\% percentiles of the distributions. The noAGN satellite population is slightly bluer starting from $z=2$, reflecting a more recent star formation, possibly impeded in the wAGN simulation by the presence of an AGN in the central galaxy.

\subsubsection{Spatial extent of the quenching}

Fig.~\ref{quiescentgas_vs_distance_various_z} illustrates the quiescent and gas poor fractions of the galaxies around the central as a function of the normalized distance to the central (the distance to the central divided by the virial radius of the central). The upper panels show the evolution of the quiescent fraction in the wAGN and noAGN simulations, and the lower panels show the evolution of the gas poor fraction with thresholds for being quiescent and gas poor identical to Fig.~\ref{quiescentgas_vs_z}. The error bars indicate the $99.7$\% confidence intervals computed for a beta distribution. The quiescent and gas poor fractions are similar at $z=3$ and the difference grows at lower redshifts, first out to twice the virial radius of the central galaxy at $z=2$, then out to more than five times the virial radius of the central galaxy at $z=1$. One caveat is the fact that in the wAGN simulations, the satellite galaxies can be affected by the AGN of other neighboring galaxies than the central, which mitigates the conclusion as to the extent of the quenching mechanism, but still points to the importance of the effect of AGN feedback from BH hosted by other galaxies.

\subsection{Histories of the main progenitors of the satellites}
\label{histories}

\begin{figure*}
    \centering
    \includegraphics[width=\textwidth]{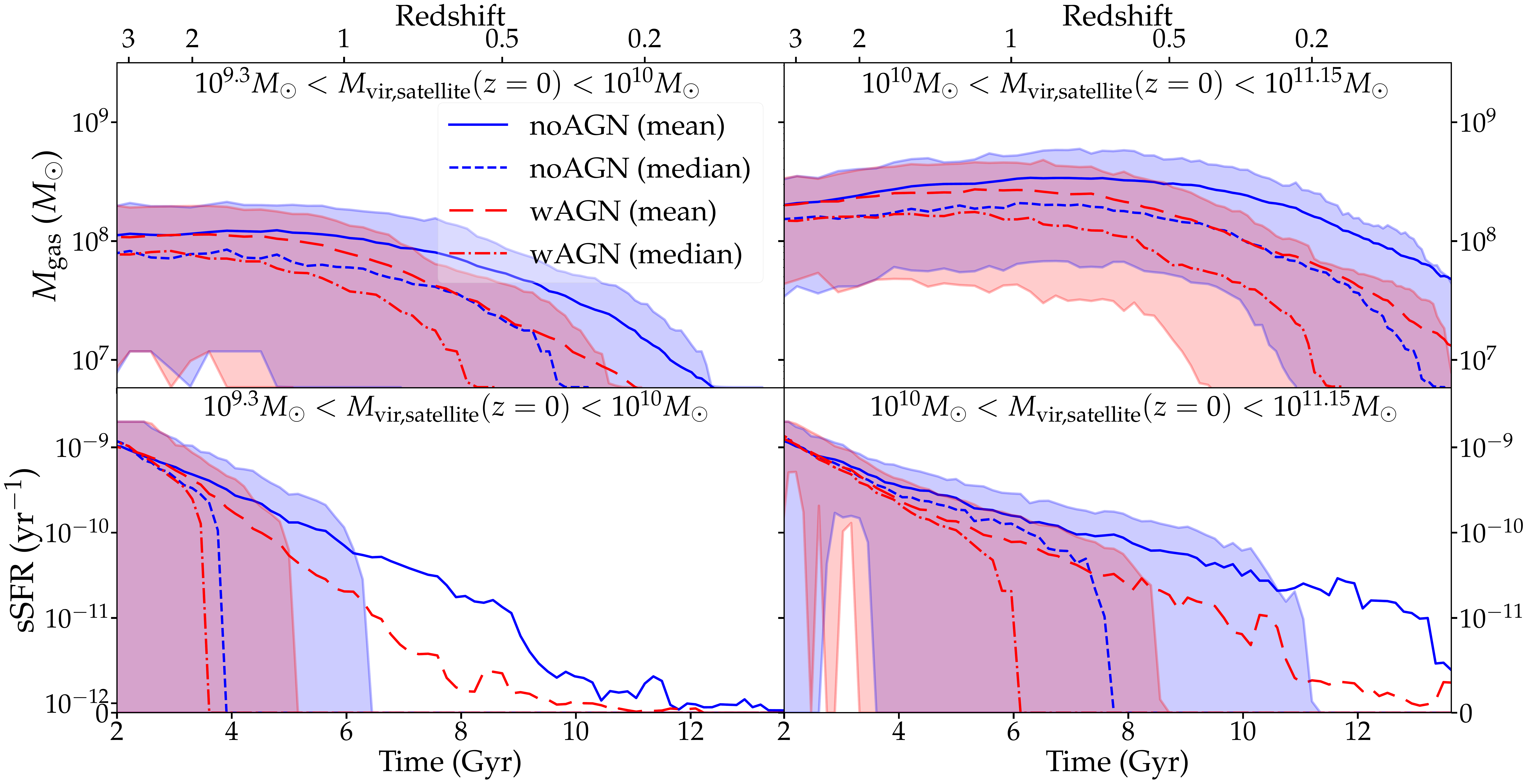}
    \caption{History of the gas mass (top) and sSFR (bottom) of the galaxies that lie inside twice the virial radius of the central galaxy at $z=0$ and traced back in time from $z=0$, in two different mass bins. The satellites are distributed in the different mass bins as a function of their virial mass at $z=0$. We stacked the satellites from the 27 zoom in simulations. The different panels show increasing mass bins as indicated at the top of the panels and contain $\sim 1400$ satellites for the lower mass bin and $\sim 350$ satellites for the higher mass bin. The shaded area shows the 16\% and the 84\% percentiles of the data. We also show the mean and the median values of the distributions.}
    \label{ssfrgas}
\end{figure*}

\begin{figure*}
    \centering
    \includegraphics[width=\textwidth]{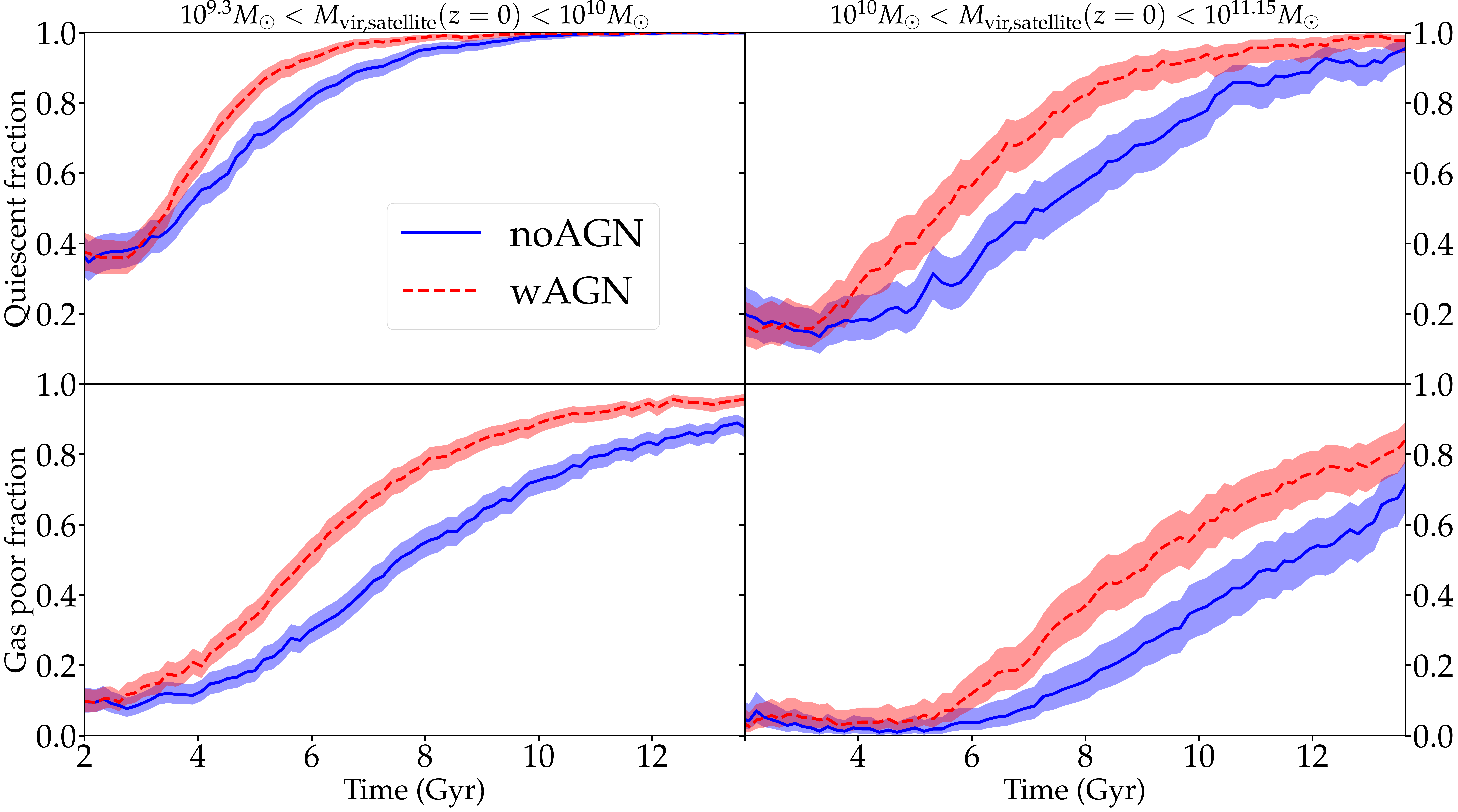}
    \caption{History of the quiescent and gas poor fractions of the galaxies that lie inside twice the virial radius of the central galaxy at $z=0$ and traced back in time from $z=0$, in two different mass bins. The satellites are distributed in the different mass bins as a function of their virial mass at $z=0$. We stacked the satellites from the 27 zoom in simulations. \textbf{Top} : Quiescent fraction (sSFR $< 0.3/t_{\rm H}$, where $t_{\rm H}$ is the age of the universe at each redshift) as a function of time, from left to right the different panels show increasing mass bins as indicated at the top of the upper panels. \textbf{Bottom}: Gas poor fractions ($M_{\rm gas}/(M_{\rm stars} + M_{\rm gas}) < 0.1 $) as a function of time. From left to right, the panels show increasing mass bins as indicated at the top of the upper panels and contain $\sim 1400$ satellites for the lower mass bin and $\sim 350$ satellites for the higher mass bin. }
    \label{fractions_histories_time}
\end{figure*}

\begin{figure*}
    \centering
    \includegraphics[width=\textwidth]{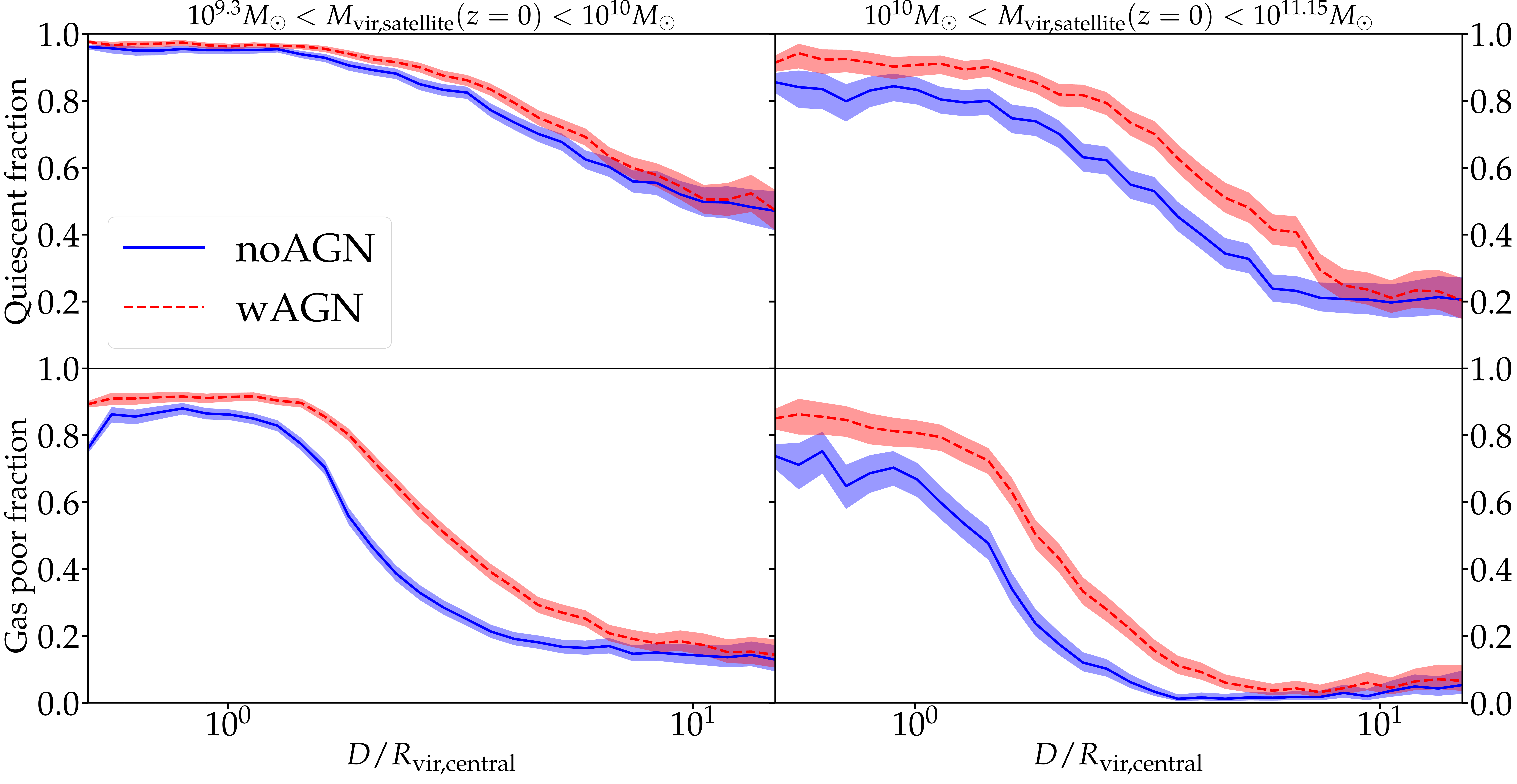}
    \caption{Quiescent and gas poor fractions of the galaxies that lie inside twice the virial radius of the central galaxy at $z=0$ and traced back in time from $z=0$, in different mass bins, as a function of their distance to the central galaxy across cosmic time, normalized to the virial radius. The satellites are distributed in the different mass bins as a function of their virial mass at $z=0$. \textbf{Top}: Quiescent fraction (sSFR $< 0.3/t_{\rm H}$, where $t_{\rm H}$ is the age of the universe at each redshift) as a function of distance to the central galaxy. \textbf{Bottom}: Gas poor fractions ($M_{\rm gas}/(M_{\rm stars} + M_{\rm gas}) < 0.1 $) as a function of time. From left to right, the panels panels show increasing mass bins as indicated at the top of the upper panels and contain $\sim 1400$ satellites for the lower mass bin and $\sim 350$ satellites for the higher mass bin.}
    \label{fractions_histories_distance}
\end{figure*}

To gain more insight into the onset and time evolution of the quenching from the central AGN, we select, in each zoom halo, the neighboring halos that are within a sphere of twice the viral radius of the central galaxy at redshift 0 and trace back in time the main progenitor (\textit{i.e.} the most massive progenitor) of these galaxies. This approach is different from the one used in section \ref{fixed z} where we select different satellites at each redshift based on a distance criterion. Here, we spatially select the galaxies at redshift zero and follow their evolution across cosmic time. Specifically, we follow their gas content and star formation history, as well as their distance to the central galaxy. As previously, we use a lower mass cut of 64 dark matter particles for selecting the satellites, and we stop tracking back the satellites when their mass is below that threshold. In both wAGN and noAGN simulations, taking into account all zoom-in simulations, there are in total $\sim 1700$ galaxies within $2 R_{\rm vir,central}$ at redshift 0.

Fig.~\ref{ssfrgas} displays the stacked evolution of the gas mass and specific star formation rate (sSFR) of the main progenitor of the neighboring galaxies at $z=0$ in two different mass bins. In each bin of mass, the presence of an AGN in the central galaxy seems to alter the gas content and star formation of the neighboring galaxies. The 84\% percentile, median and mean values of the gas mass distributions are lower for the wAGN simulations, perhaps suggesting a slightly reduced accretion onto the galaxies. The decrease of the gas content occurs earlier in the wAGN simulation than in the noAGN simulation, indicating that the central AGN is involved in direct gas removal from the neighboring galaxies, or indirect removal through starvation. The quenching of star formation in the satellite galaxies also takes place earlier in the wAGN simulation than in the noAGN simulation.

Fig.~\ref{fractions_histories_time} shows, as a function of cosmic time, the stacked evolution of the quiescent and gas poor fractions of the same traced back sample, in different mass bins. The difference in the histories in the two wAGN and noAGN simulations develops at cosmic times of 4 to 8 Gyr, depending on the mass bin. The difference between noAGN and wAGN in gas poor and quiescent satellite fractions peaks at times of 6 to 10 Gyr depending on the mass bin.

Fig.~\ref{fractions_histories_distance} shows the same evolution as Fig.~\ref{fractions_histories_time} but as a function of the distance to the central galaxy instead of cosmic time. We normalize that distance to the virial radius of the central galaxy. Note that Fig.~\ref{fractions_histories_distance} is not a profile plot of the quiescent and gas poor fractions as a function of distance to the central at a given time: for each distance bin, the contribution of each galaxy to the distribution comes from different cosmic times. Note also that the distance plotted on the $x$ axis is normalized to the virial radius of the central at the time when the distance is computed, which means that the physical value of that radius grows as a function of time, and that a given normalized distance is smaller in physical units at higher redshifts than lower redshifts because of the growth of the virial radius of the central galaxy. Fig.~\ref{fractions_histories_distance} shows that the normalized distance below which the wAGN and noAGN simulations differ is of 5 to 10 times the virial radius of the central galaxy. This means that the onset of the quenching by the central AGN occurs as early as when the galaxy is at a distance from 5 to 10 times the virial radius of the central. As mentioned before, a caveat is that in the wAGN simulation, throughout their journey towards the central galaxy, satellite galaxies can also be affected by the AGN of other massive galaxies than the central.

\section{Quenching mechanism}
\label{physics}

\begin{figure*}
  \centering
  \subfloat{\includegraphics[width=\textwidth]{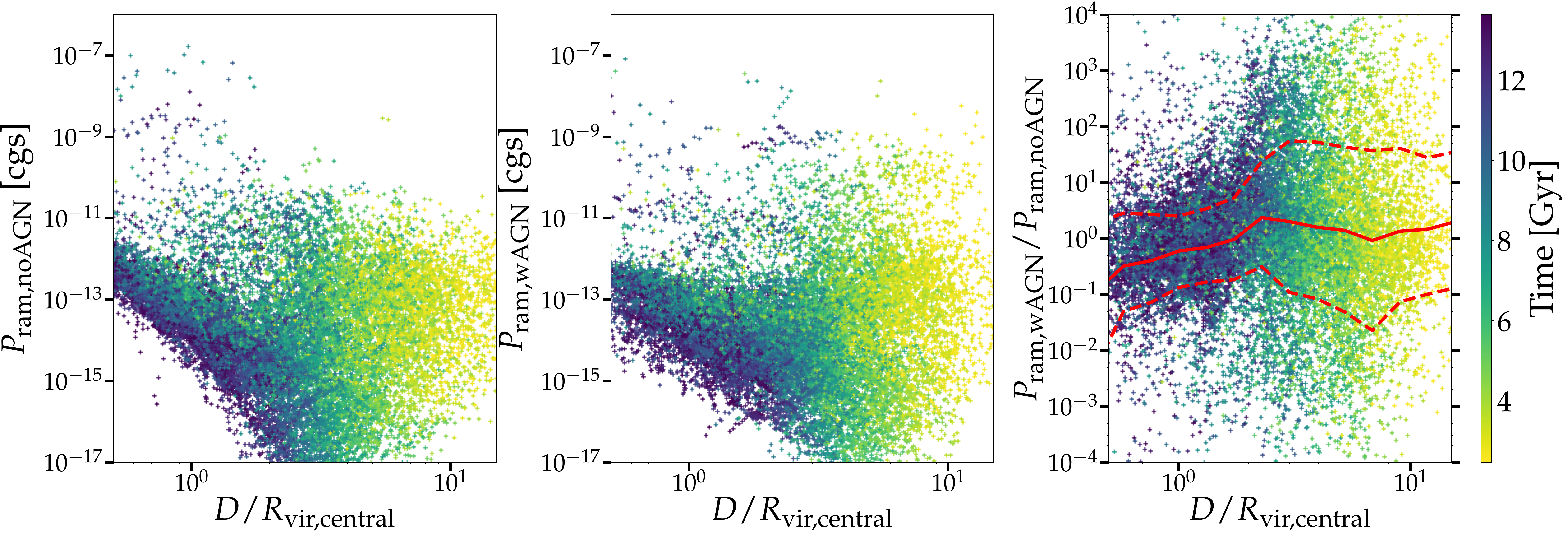}}
  \hspace{5pt}
  \subfloat{\includegraphics[width=\textwidth]{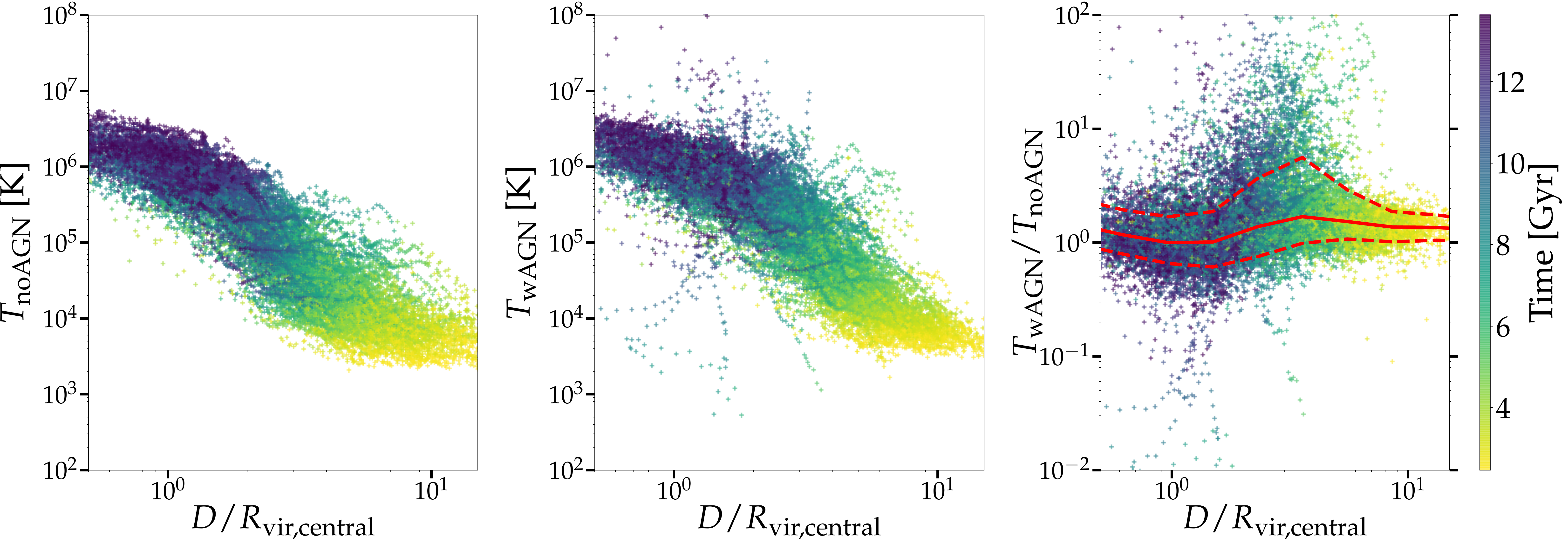}}
  \hspace{5pt}
  \subfloat{\includegraphics[width=\textwidth]{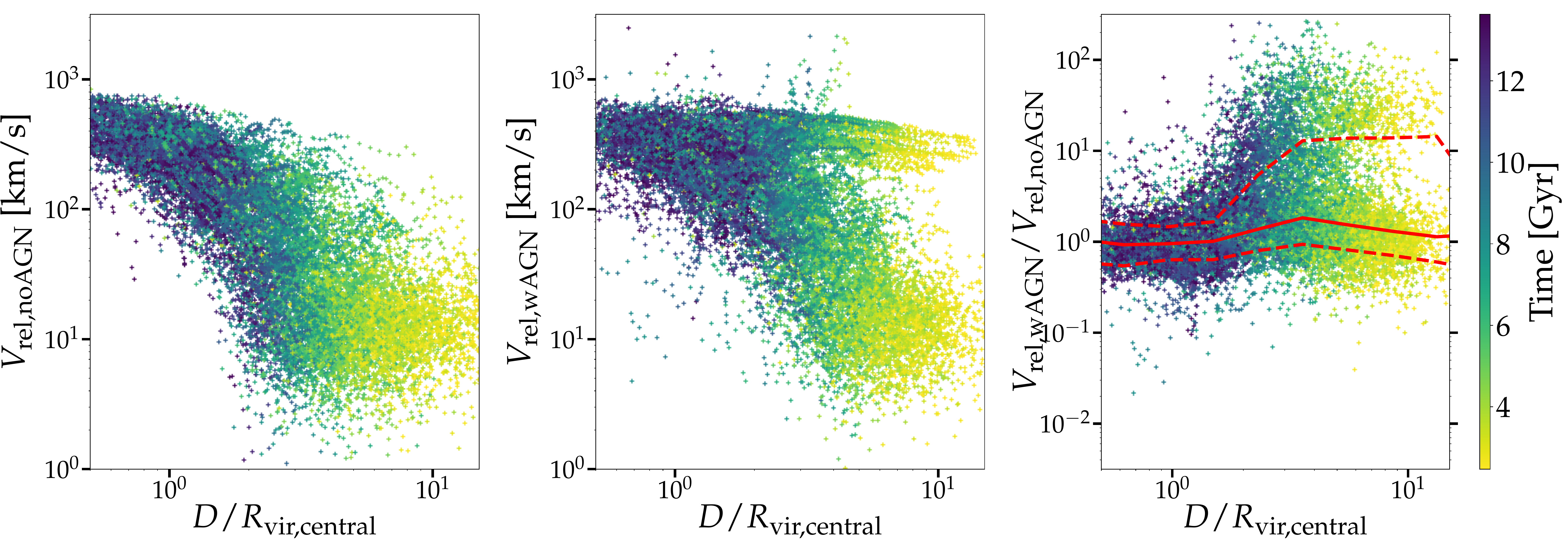}}
  \caption{Physical conditions in a shell around the main progenitors of the galaxies that lie inside twice the virial radius of the central at $z=0$. As previously, we use a lower mass cut of 64 dark matter particles for selecting the satellites, and we stop tracking back the satellites when their mass is below that threshold. We also exclude from the noAGN and wAGN sample the satellites that are above the mass threshold for BH seeding. In order to compare the wAGN and noAGN simulations, we only show satellites that are matched in pairs across the wAGN and noAGN simulations, which is the case if 50\% of their dark matter particles have identical identifiers at $z=0$. That criterion is fulfilled by $\sim 260$ pairs. All the physical quantities are mass averaged in a shell around the satellite between $0.1 R_{\rm vir,satellite}$ and $2 R_{\rm vir,satellite}$. Each dot represents the physical quantity around one satellite at one snapshot, as a function of the distance to the central normalized to the virial radius of the central. Each row contains three panels: the first panel displays the physical conditions in the noAGN simulation, the second in the wAGN simulation and the third displays the ratio between both. \textbf{Upper panels:} Ram pressure undergone by the satellite, \textit{i.e.} the product $\rho V_{\rm rel}^2$ where $\rho$ is the mass averaged gas density around the satellite and $V_{\rm rel}$ the gas velocity around the satellite relative to the velocity of the satellite. \textbf{Middle panels:} Mass averaged temperature around the satellite galaxies. \textbf{Lower panels:} mass averaged gas velocity relative to the satellite velocity. The solid red lines indicate the median values, the dashed lines correspond to the 16\% and the 84\% percentiles of the distributions. The dots are color-coded with the corresponding cosmic time.}
  \label{physicalconditions}
\end{figure*}

\begin{figure*}
  \subfloat{\includegraphics[width=\textwidth]{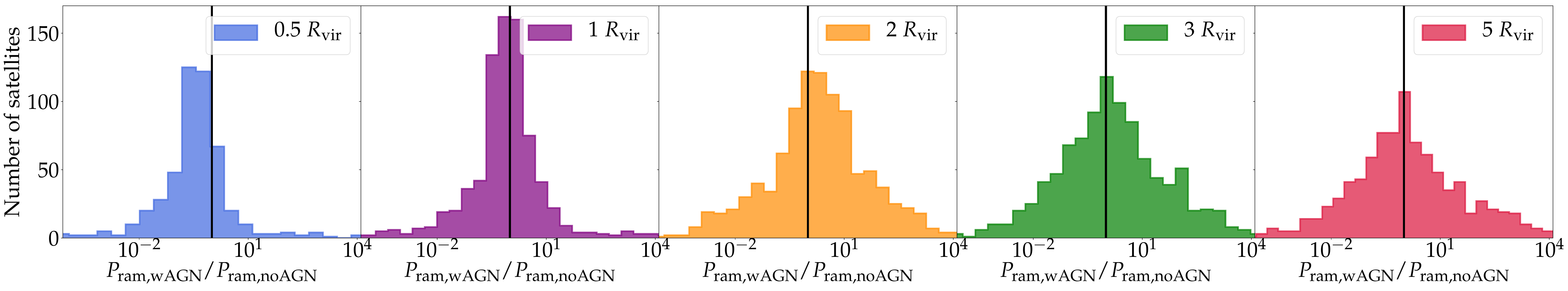}}
        \hspace{5pt}
  \subfloat{\includegraphics[width=\textwidth]{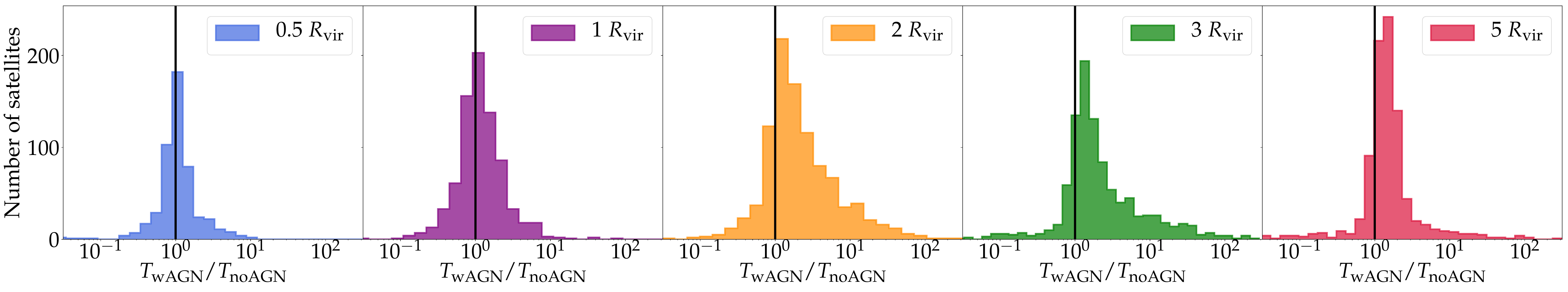}}
        \hspace{5pt}
    \subfloat{\includegraphics[width=\textwidth]{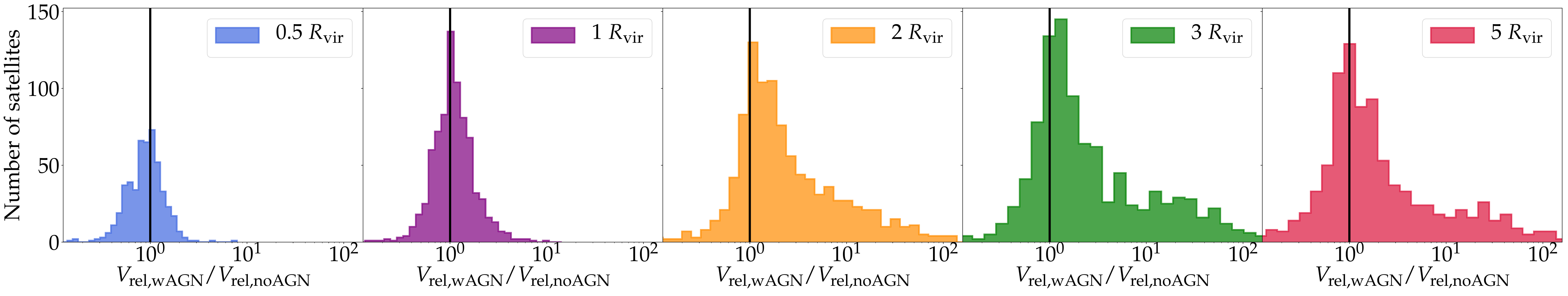}}
   \caption{Histograms of the ratio of physical quantities around pairs of wAGN and noAGN satellites. The quantities are computed in a shell around the main progenitors of the galaxies that lie inside twice the virial radius of the central at $z=0$. As previously, we use a lower mass cut of 64 dark matter particles for selecting the satellites, and we stop tracking back the satellites when their mass is below that threshold. We also exclude from the noAGN and wAGN sample the satellites that are above the mass threshold for BH seeding. In order to compare the wAGN and noAGN simulations, we only show satellites that are matched in pairs across the wAGN and noAGN simulations, which is the case if 30\% of their dark matter particles have identical identifiers at $z=0$. That criterion is fulfilled by $\sim 1000$ pairs. All the physical quantities are mass averaged in a shell around the satellite between $0.1 R_{\rm vir,satellite}$ and $2 R_{\rm vir,satellite}$. The three rows show the ratios $P_{\rm ram,wAGN}/ P_{\rm ram,noAGN}$ (top row), $T_{\rm wAGN}/ T_{\rm noAGN}$ (middle row),  $V_{\rm rel,wAGN}/ V_{\rm rel,noAGN}$ (bottom row). Each row contains five histograms: the ratios at $0.5,\,1,\,2,\,3,\,5\, R_{\rm vir,central}$ that the satellites experience during their trajectory towards the central. The ram pressure undergone by the satellite, temperature and relative velocity of the gas are computed in the same way as in Fig.~\ref{physicalconditions}}
  \label{histograms}
\end{figure*}

\begin{figure*}
  \centering
 \includegraphics[width=\textwidth]{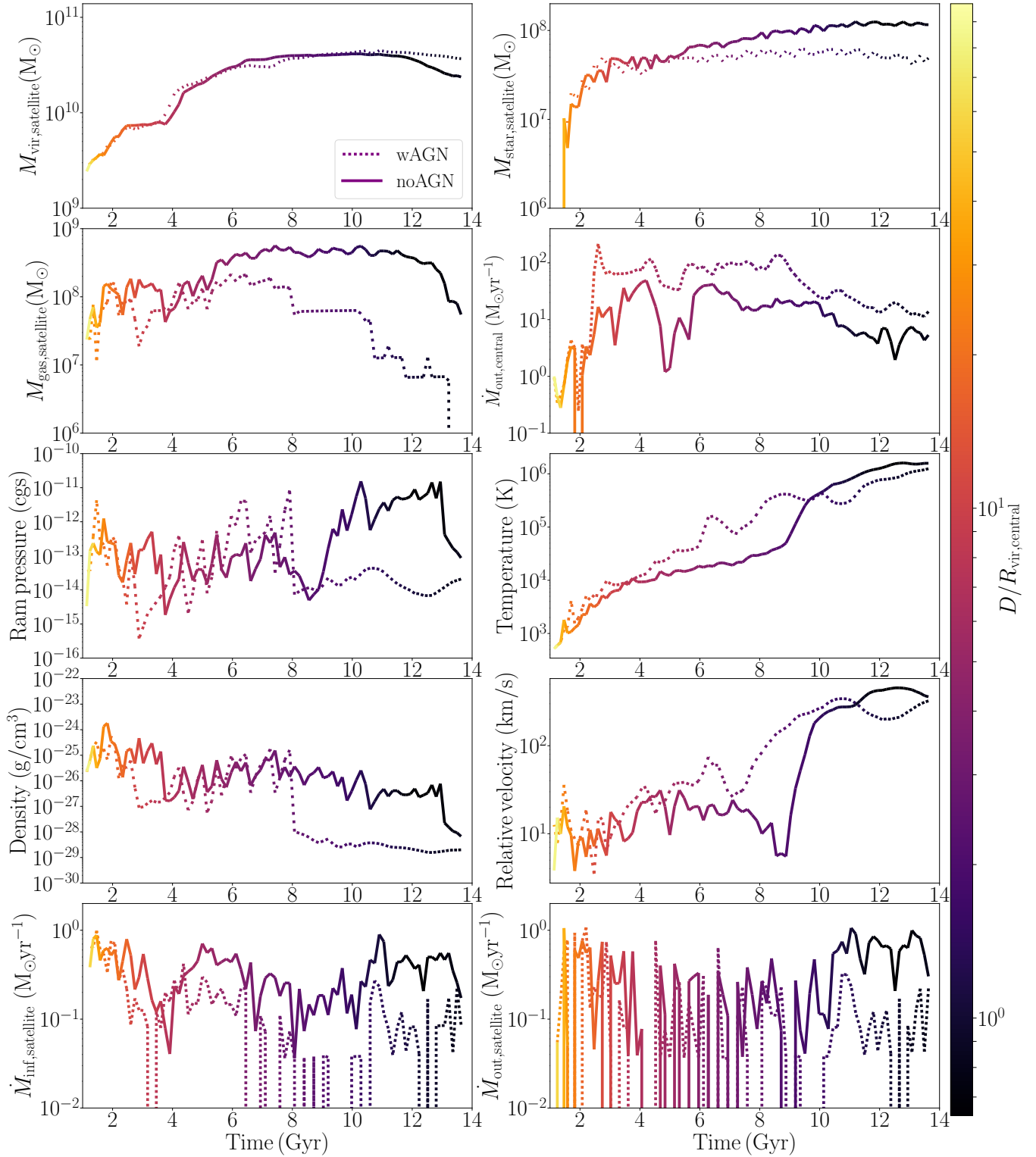}
 \caption{For one satellite in particular, traced back in the noAGN (solid line) and wAGN simulations (dashed line), physical quantities around the satellite are plotted as a function of time. The stellar, gaseous and virial masses of the satellite are also plotted, as well as the mass outflow rate from the central at one virial radius from the central. The inflow rate onto and outflow rate from the satellite are computed at $ 0.3 R_{\rm vir,satellite}$ and plotted in the bottom panels. The density, relative velocity of the gas and temperature are mass averaged in a shell around the satellite between $0.1 R_{\rm vir,satellite}$ and $2 R_{\rm vir,satellite}$. The ram pressure is the product $\rho V_{\rm rel}^2$ where $\rho$ is the mass averaged density around the satellite and $V_{\rm rel}$ the gas velocity around the satellite relative to the velocity of the satellite. The lines are color-coded with the distance to the central normalized to the virial radius of the central. The virial mass and radius of the central galaxy are close to respectively $8 \times 10^{12}\, \rm{M}_{\odot}$ and $500$ kpc at redshift 0. }
 \label{215}
  \end{figure*}

\begin{figure*}
  \subfloat[wAGN: z=2.13]{\includegraphics[width=0.3\textwidth]{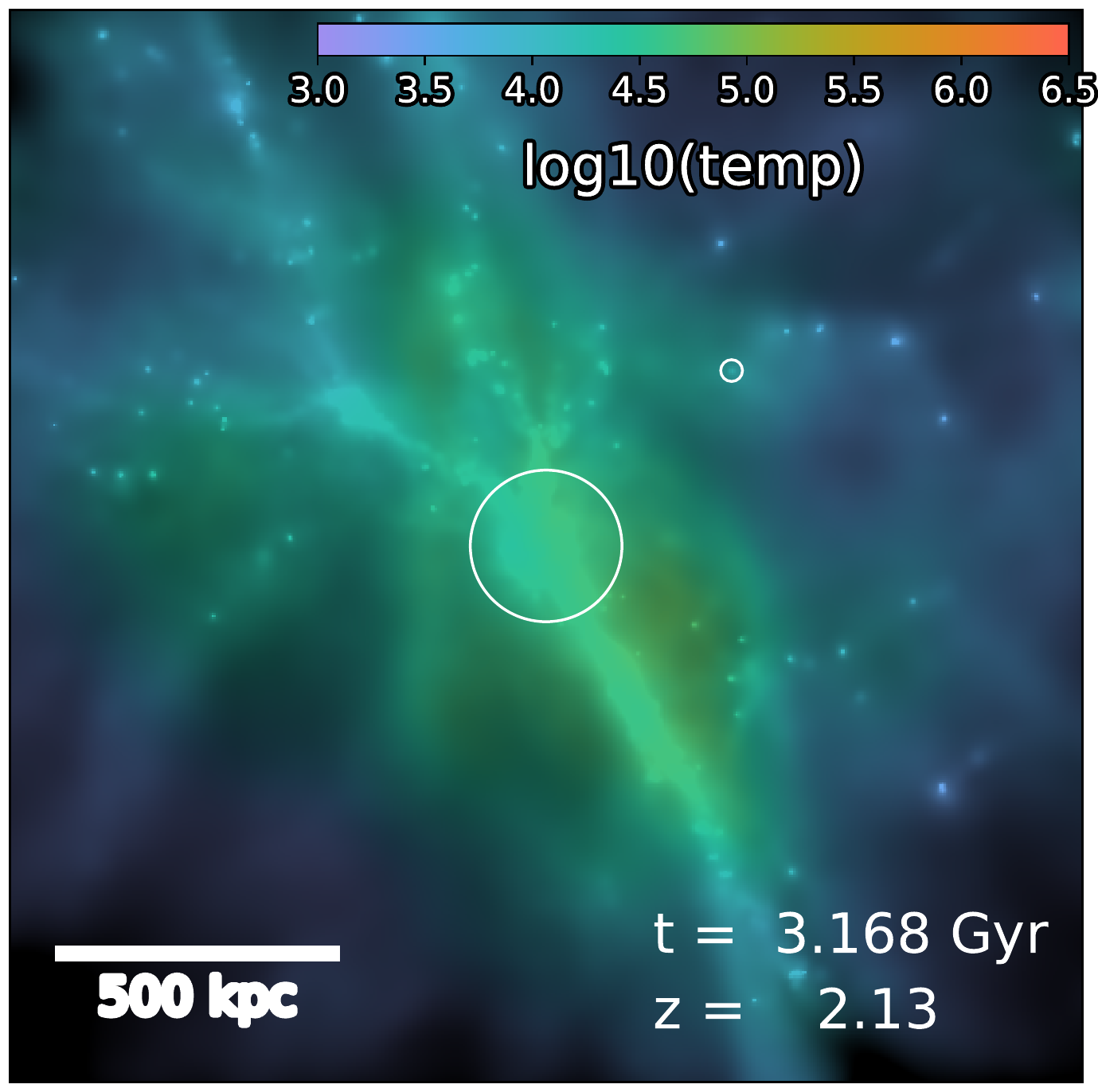}}
  \subfloat[wAGN: z=1.08]{\includegraphics[width=0.3\textwidth]{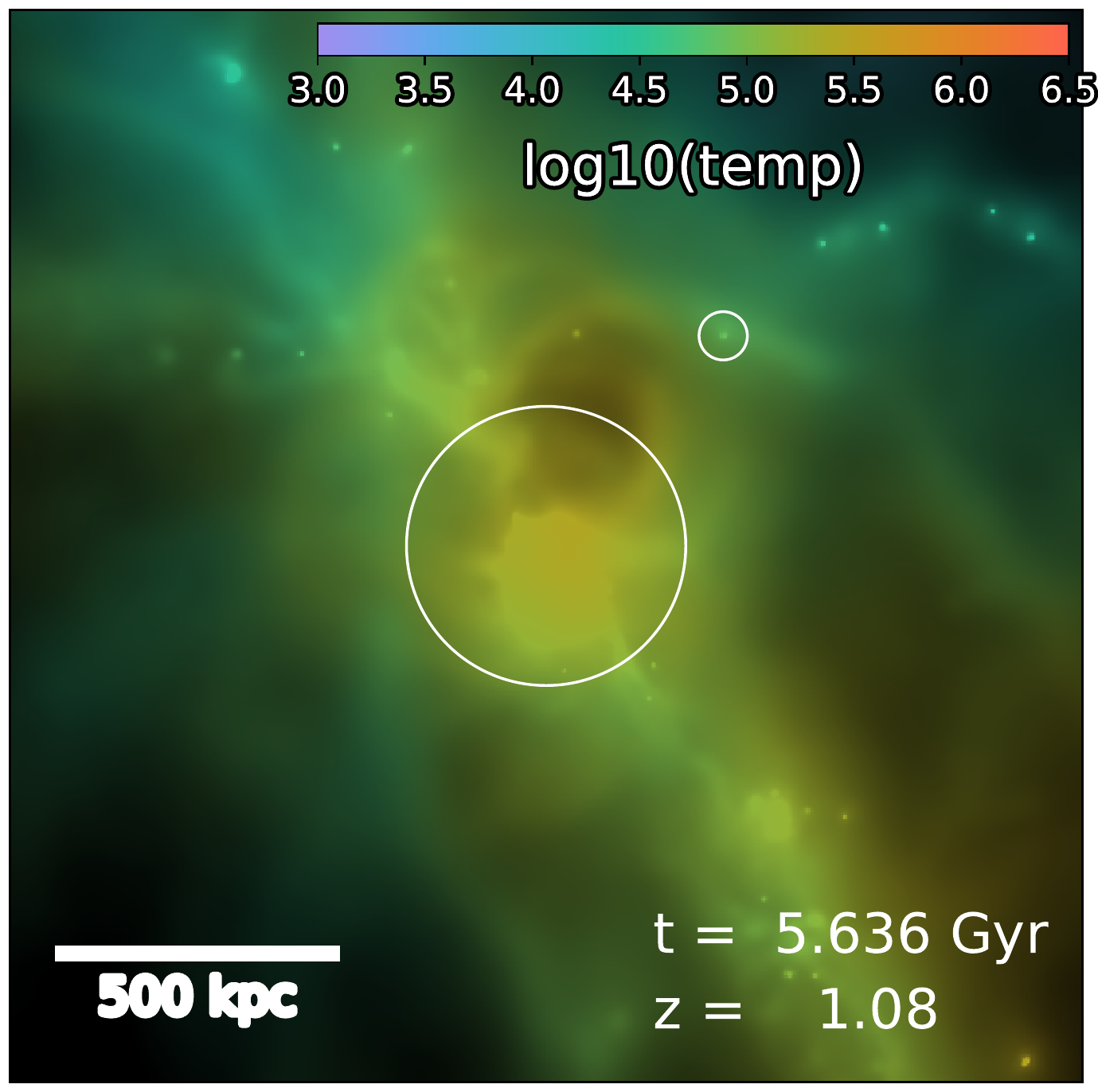}}
    \subfloat[wAGN: z=0.45]{\includegraphics[width=0.3\textwidth]{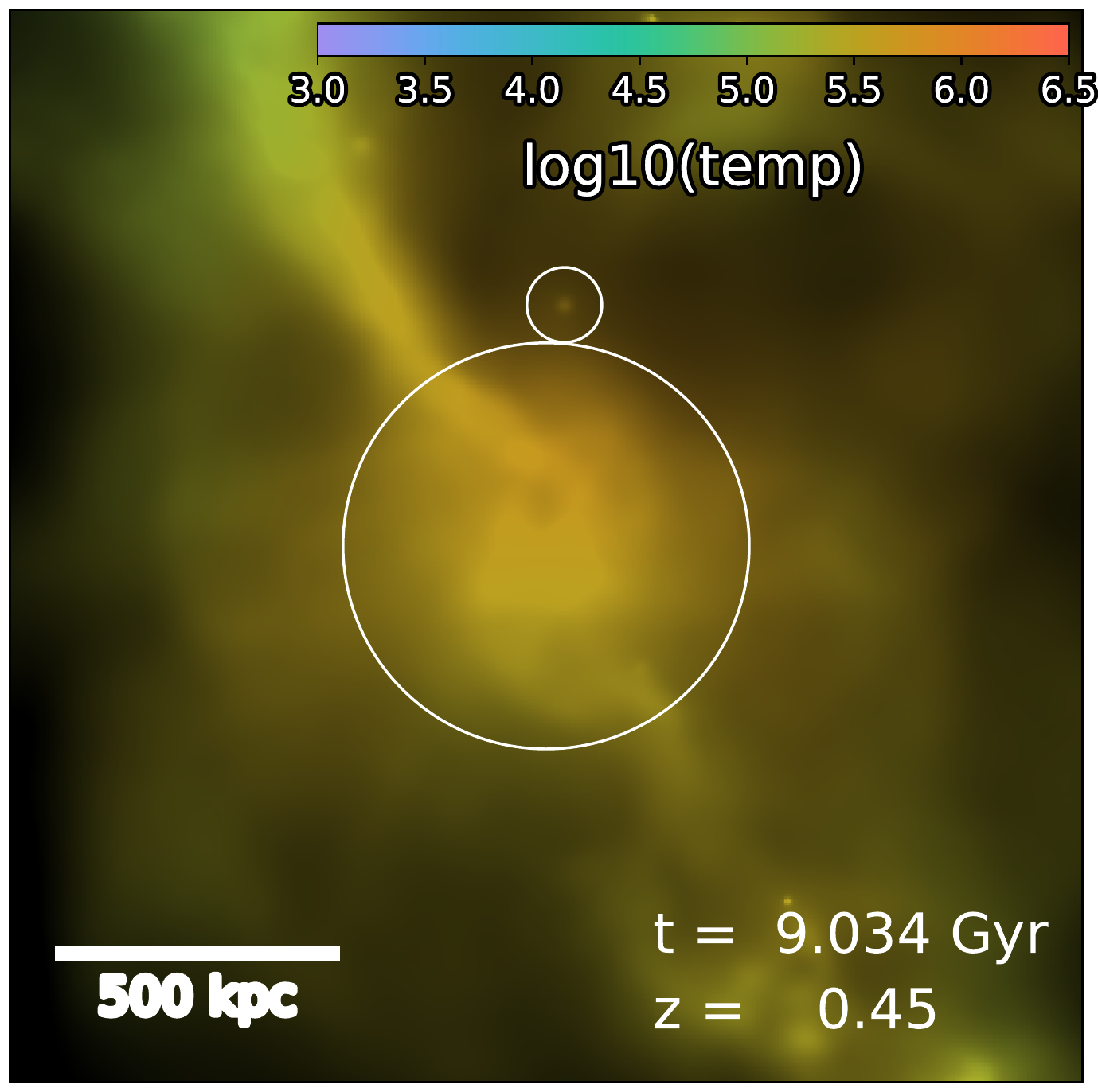}}
      \hspace{5pt}
  \subfloat[noAGN: z=2.13]{\includegraphics[width=0.3\textwidth]{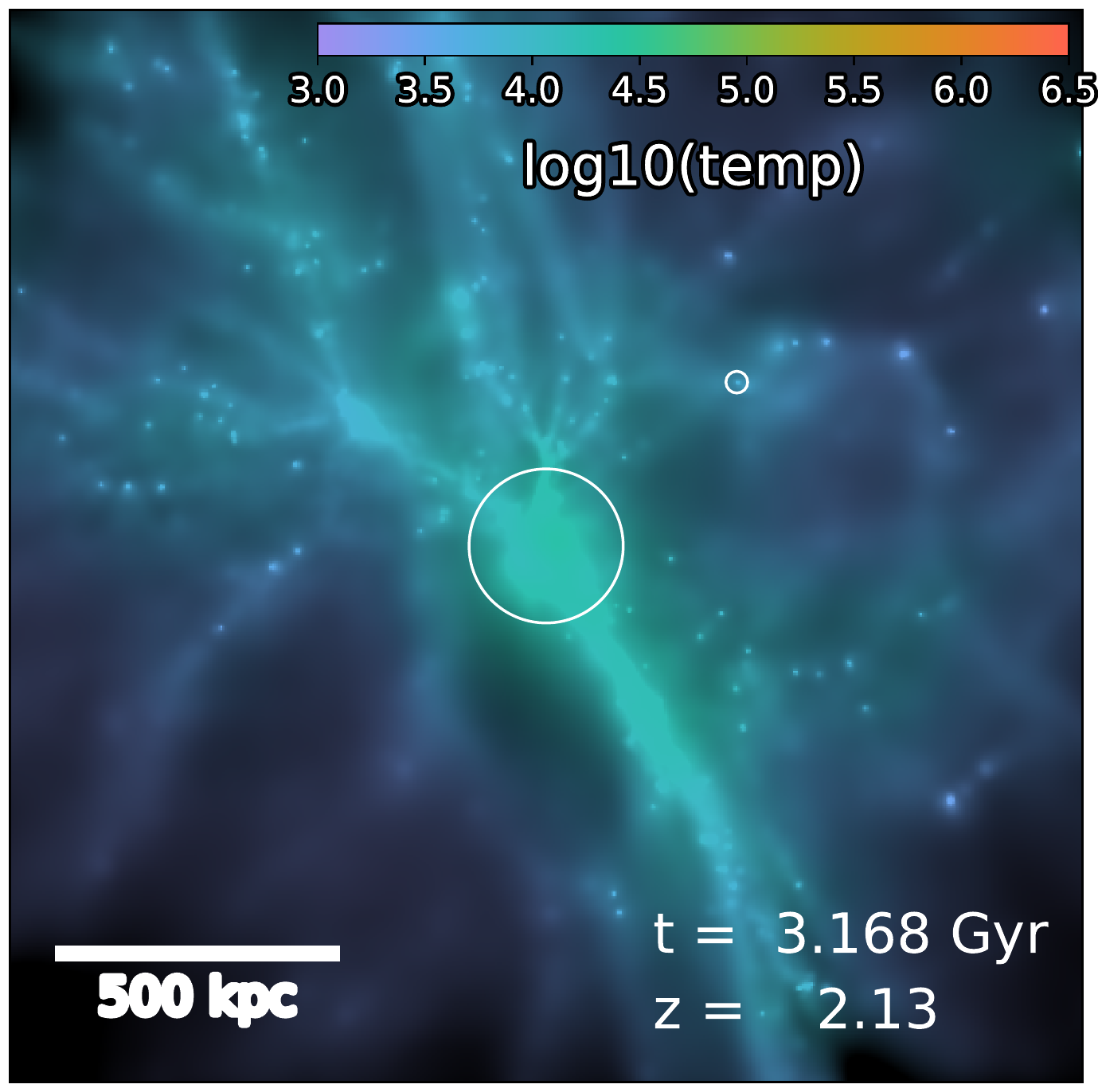}}
  \subfloat[noAGN: z=1.08]{\includegraphics[width=0.3\textwidth]{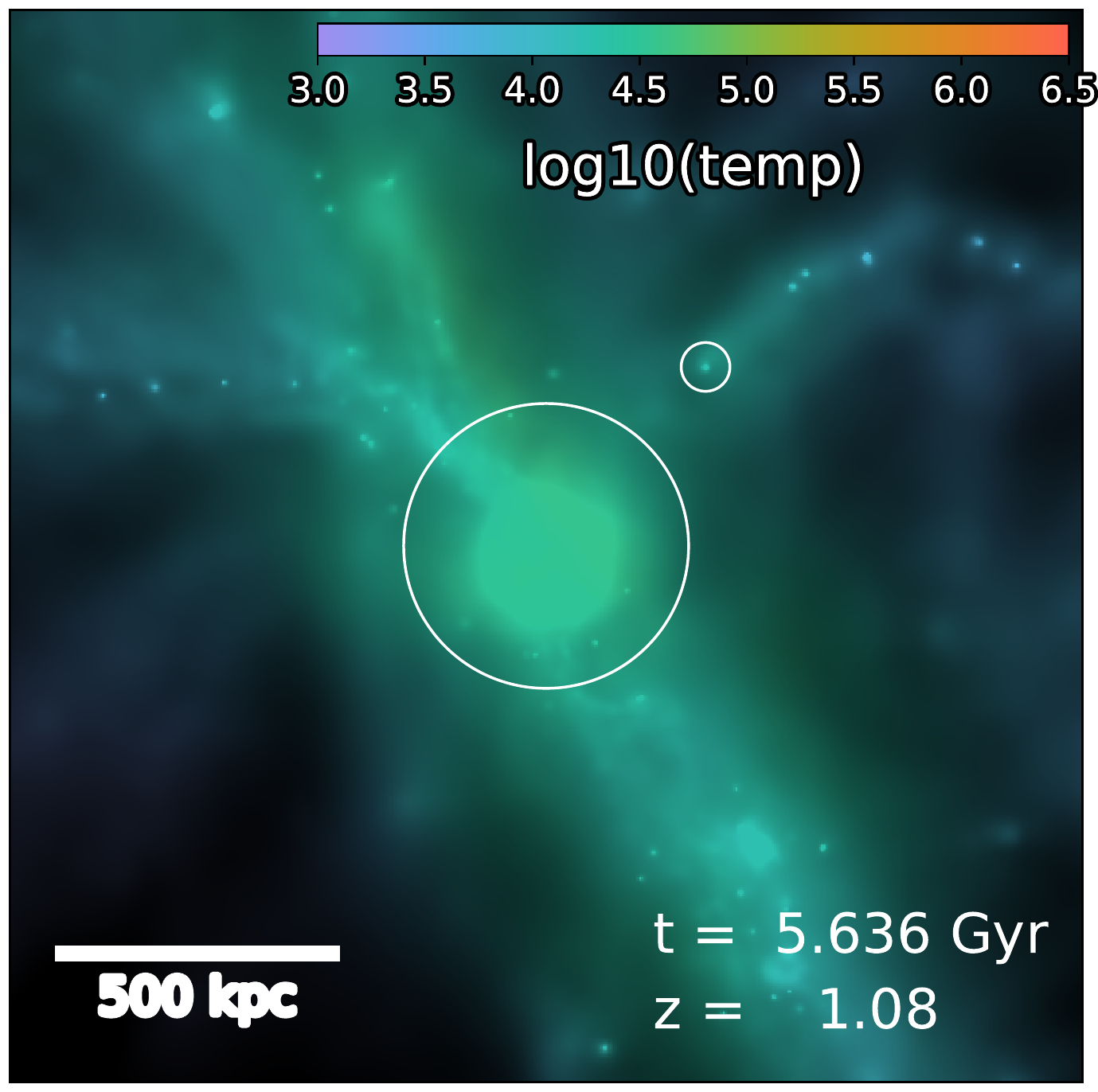}}
    \subfloat[noAGN: z=0.45]{\includegraphics[width=0.3\textwidth]{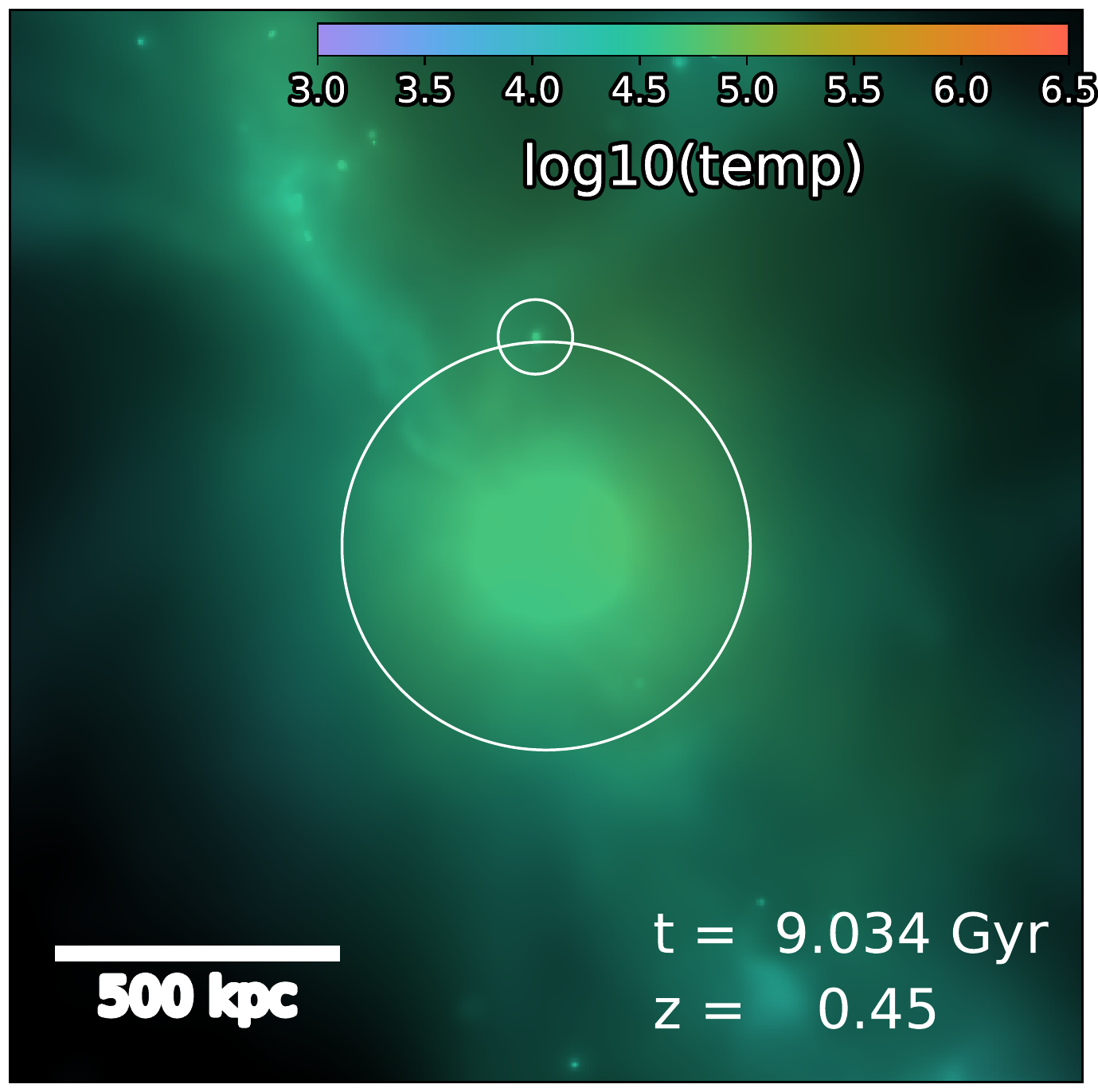}}
  \caption{The satellite analyzed in Fig.~\ref{215} is shown at different cosmic times, orbiting the central galaxy shown at the center of each frame. The white circles indicate the virial radius of each galaxy. The three upper panels show three snapshots in the wAGN simulation. The three lower panels show the same snapshots in the noAGN simulation. The brightness shows the surface density and the color traces the temperature. The virial mass and radius of the central galaxy are close to respectively $8 \times 10^{12} \rm{M}_{\odot}$ and $500$ kpc at redshift 0.}
  \label{movie}
\end{figure*}

 In this section, we investigate the nature of the physical mechanism causing the additional quenching of the satellites in the wAGN simulation. We study the effect of the wind from the central AGN by following the ram pressure undergone by the satellite galaxies, the temperature in their surroundings and relative velocity of the intergalactic gas. We compute these quantities in a shell around the satellite, between $0.1 R_{\rm vir,satellite}$ and $2 R_{\rm vir,satellite}$. Fig.~\ref{physicalconditions} displays these physical quantities for satellite galaxies throughout the simulation. We trace back in time the main progenitors of the galaxies that lie inside twice the virial radius of the central at $z=0$. As previously, we use a lower mass cut of 64 dark matter particles for selecting the satellites, and we stop tracking back the satellites when their mass is below that threshold. In order to compare the wAGN and noAGN simulations, we only show satellites that are matched in pairs across the wAGN and noAGN simulations, which is the case if 50\% of their dark matter particles have identical identifiers at $z=0$. That criterion is fulfilled by $\sim 260$ pairs which corresponds to a fraction of $\sim 18\%$ of the satellite galaxies that lie within $2R_{\rm vir,central}$ at $z=0$. Each dot represents the physical quantity around one satellite galaxy at one snapshot, as a function of the distance to the central normalized to the virial radius of the central. 
 
The ram pressure undergone by the satellite is the product $P_{\rm ram}=\rho V_{\rm rel}^2$ where $\rho$ is the mass averaged gas density around the satellite and $V_{\rm rel}$ the gas velocity around the satellite relative to the satellite.  The ram pressure, temperature and relative velocity of the gas around the satellite are affected by AGN feedback throughout their trajectory toward the central, mostly at a distance of two to three times the virial radius from the central. At a distance equal to three times the virial radius of the central, the 16\% and 86\% percentiles are around 0.2 and 60 respectively, for the ratio of ram pressures $P_{\rm ram,wAGN}/P_{\rm ram,noAGN}$; 1 and 6 respectively, for the ratio of temperatures $T_{\rm wAGN}/T_{\rm noAGN}$, 0.8 and 11 respectively, for $V_{\rm rel,wAGN}/V_{\rm rel,noAGN}$. This hints at possible quenching mechanisms to explain the additional suppression of star formation and gas loss of the satellites in the wAGN simulations. The ram pressure from the wind driven by the AGN can directly strip the gas, in a manner similar to \textit{ram pressure stripping} where the velocity is not only the orbital velocity of the satellite, but also that of the wind coming from the central AGN. Moreover, the higher temperatures and velocity of the intergalactic gas can also prevent the gas from accreting efficiently onto the satellites, in a manner similar to \textit{strangulation}.

Fig.~\ref{histograms} shows the histograms of the ratios $P_{\rm ram,wAGN}/ P_{\rm ram,noAGN}$, $T_{\rm wAGN}/ T_{\rm noAGN}$,  $V_{\rm rel,wAGN}/ V_{\rm rel,noAGN}$ at $0.5,\,1,\,2,\,3,\,5\, R_{\rm vir,central}$ that the satellites experience during their trajectory towards the central. The sample that we chose is the following: amongst the main progenitors of the satellites that lie within $2 R_{\rm vir,central}$ at $z=0$, we show satellites that are matched in pairs across the wAGN and noAGN simulations (when 30\% of their dark matter particles have identical identifiers at $z=0$; note that this threshold of $30\%$ is different that the one used in Fig.~\ref{physicalconditions}, where it was higher for the clarity of the plot). That criterion is fulfilled by $\sim 1000$ pairs.\\
The distribution of the ratios of temperatures and relative velocities shows that beyond $2 R_{\rm vir,central}$, an important fraction of the satellites in the wAGN simulation experience higher temperatures and relative velocities of the gas in their surroundings than the satellites in the noAGN simulation. This shows that the physical origin of the environmental influence of the central AGN extends as far as 3 to 5 times the virial radius of the central.
However, the differences in ram pressure are less conclusive, which suggests that it is primarily the increase of temperature and relative velocity of the gas in the surroundings of the  wAGN satellites that causes the differences in the evolution of the stellar and gaseous mass of the satellites in the wAGN simulation.
 
In Fig.~\ref{215} we study in detail an example of pairs of wAGN/noAGN satellite galaxies. Fig.~\ref{movie} displays three snapshots showing the evolution of the same satellite. The outflow rate from the central is computed at one virial radius from the central. The increase of the outflow rate from the central in the wAGN simulation is likely due to the central AGN. The first peak of the ram pressure, temperature and relative velocity of the gas for the wAGN satellite corresponds to the time ($\sim$ 6 Gyr) when the gaseous and stellar masses of the wAGN satellite diverge from the gaseous and stellar masses of the noAGN satellite: at that time the satellite is at a few virial radii from the central, which is why the first peak of ram pressure is observed with some delay with respect to the first peak of the outflow rate from the central that is measured at $R_{\rm vir,central}$. The panel (b) in Fig.~\ref{movie} shows the satellite orbiting the central before the passage of the wind that most likely triggers the increase of ram pressure, temperature and relative velocity of the gas at 6 Gyr in the satellite galaxy. Around 9 Gyr, the increase of the ram pressure in the noAGN simulation compared to the wAGN simulation is due to the fact that towards the center, inside the virial radius of the central, the surface density of the gas is several orders of magnitude higher in the noAGN simulation than in the wAGN simulation, as measured in \cite{Choi15}. The bottom panels show the inflow rate onto and outflow rate from the satellite, computed at $0.3 R_{\rm vir,satellite}$. The inflow rate onto the satellite in the wAGN simulations is noticeably lower than in the noAGN simulations but not the outflow rate, which suggests that a deficit of accretion rather than gas ejection is responsible for the lower gas fraction in the wAGN simulation.

\begin{figure*}
    \centering
    \includegraphics[width=\textwidth]{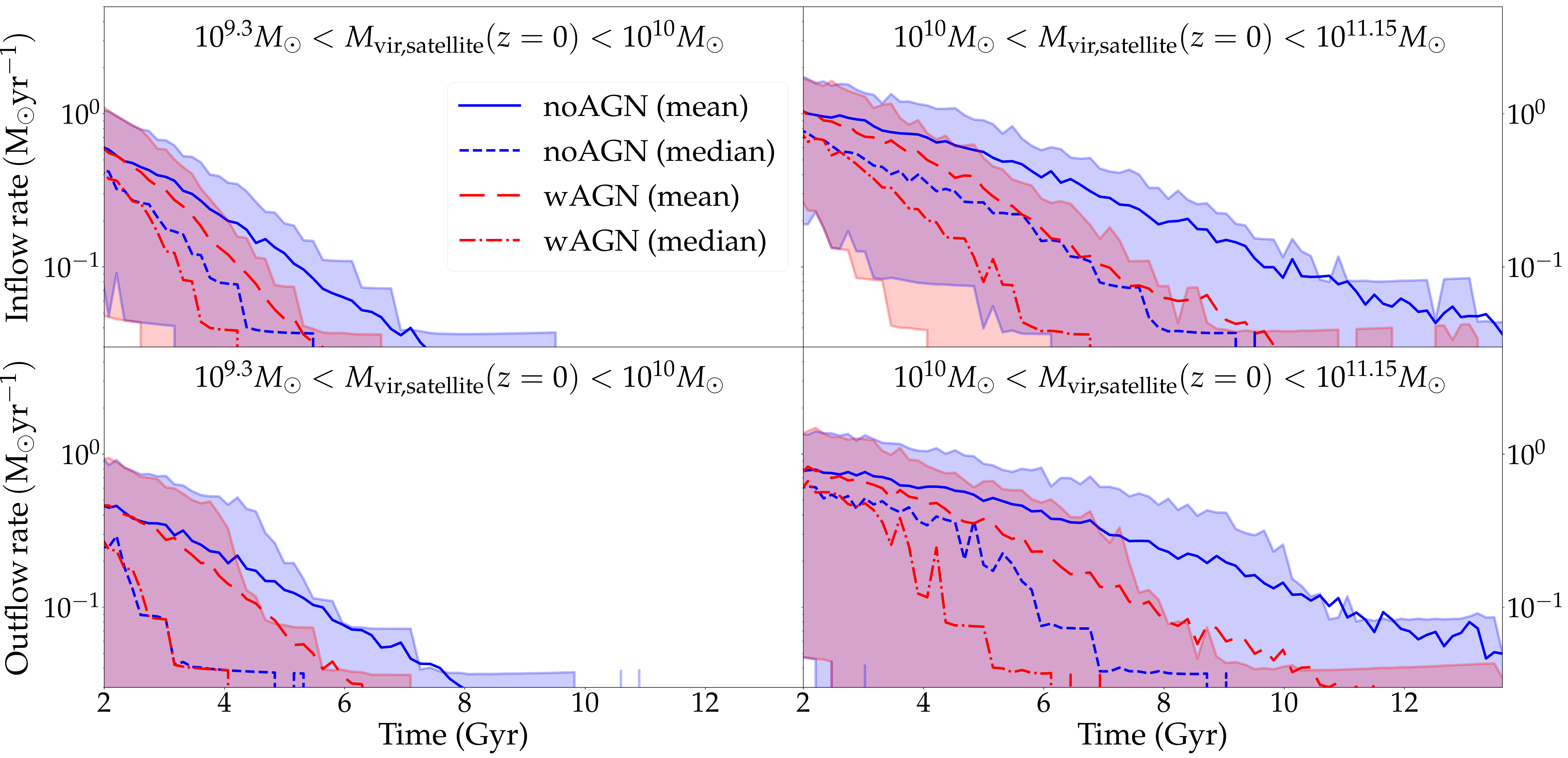}
    \caption{History of the inflow onto and outflow rate from the galaxies that lie inside twice the virial radius of the central galaxy at $z=0$ and traced back in time from $z=0$, in two different mass bins. The satellites are distributed in the different mass bins as a function of their virial mass at $z=0$. We stacked the satellites from the 27 zoom in simulations. The different panels show increasing mass bins as indicated at the top of the panels and contain $\sim 1400$ satellites for the lower mass bin and $\sim 350$ satellites for the higher mass bin. The shaded area shows the 16\% and the 84\% percentiles of the data. We also show the mean and the median values of the distributions. The inflow and outflow rates are computed at $0.1 R_{\rm vir, satellite}$.}
    \label{inflow_outflow}
\end{figure*}

Fig.~\ref{inflow_outflow} shows the history of the inflow onto and outflow rate from the galaxies that lie inside twice the virial radius of the central galaxy at $z=0$ and traced back in time from $z=0$, in two different mass bin. The inflow and outflow rates are computed at $0.1 R_{\rm vir, satellite}$. In each bin of mass, the presence of an AGN in the central galaxy seems to decrease the inflow of gas onto the neighboring galaxies. The 84\% percentile, median and mean values of the distributions are lower for the wAGN simulations, perhaps suggesting a slightly reduced accretion onto the galaxies. On the contrary, the outflow rate from the satellite is larger in the noAGN simulations. We think that it is due to the higher star formation rates and therefore stronger stellar feedback in the noAGN simulations. Overall this suggests that a deficit of accretion rather than gas ejection is responsible for the lower gas fraction in the wAGN simulation. That deficit of accretion can be due to the higher temperatures and higher velocities of the gas surrounding the satellites, as observed in Fig.~\ref{histograms}.

\section{Summary and discussion}

In this work, we explore the effects of the AGN of a central galaxy on its satellites. We used zoom-in simulations of 27 massive galaxies with halo masses of $10^{12-13.4} M_{\odot}$ at z = 0. We compared two sets of simulations: one that contains BHs and AGN feedback and one without BHs and without AGN feedback. Both simulations include stellar feedback from multiple processes, including powerful winds from SNe, stellar winds from young massive stars, and AGB stars, as well as radiative heating within Str\"omgren spheres, additional heating effects due to the presence of metals, including grain photoelectric heating and metallicity-dependent from the cosmic X-ray background. Our wAGN model is identical except that it also includes a model for BH seeding and accretion, as well as AGN feedback via high-velocity broad absorption line winds and Compton/photoionization heating. Note that our implementation of AGN feedback is different from the one implemented in most cosmological simulations, such as Illustris \citep{Illustris} and Horizon-AGN \citep{Dubois14}, in that it allows for broad absorption line winds and X-ray and UV output of both energy and momentum. We use the noAGN simulations as a control sample to isolate the effect of the central AGN. 
We find that the inclusion of AGN feedback from centrals produces a significant difference in the evolution of the gas content and the star formation in the satellite galaxies, which suggests that AGN feedback from the central BHs hosted by other galaxies plays a substantial role in the evolution of the satellite galaxy population. We also explore the spatial extent of this satellite quenching mechanism: we find that the difference between the two simulations extends as far out as five times the viral radius of the central galaxy at $z=1$, suggesting that this effect could be relevant to models trying to reproduce the observed patterns of environmental quenching and perhaps to the physical origin of \textit{galactic conformity}. We explored the physical mechanisms responsible for the additional quenching observed in the wAGN simulations and argue that the wind from the central AGN can increase the relative velocity and temperature of the gas in the surrounding of the satellites causing a lack of gas accretion onto them. A possible caveat is the fact that in the wAGN simulations, the satellite galaxies can be affected by the AGN of other satellites and not only the central galaxy. We will explore in more detail the physics of this quenching mechanism in subsequent studies using idealized simulations.

\label{Conclusion}

\section*{Acknowledgements}
We thank the anonymous referee for useful remarks. This work was initiated as a project for the Kavli Summer Program in Astrophysics held at the Center for Computational Astrophysics of the Flatiron Institute in 2018. The program was co-funded by the Kavli Foundation and the Simons Foundation. We thank them for their generous support. GD thanks Yohan Dubois, Gary Mamon and Joseph Silk for very fruitful discussions.




\bibliographystyle{mnras}
\bibliography{paper} 

\begin{thebibliography}{}
\makeatletter
\relax
\def\mn@urlcharsother{\let\do\@makeother \do\$\do\&\do\#\do\^\do\_\do\%\do\~}
\def\mn@doi{\begingroup\mn@urlcharsother \@ifnextchar [ {\mn@doi@}
  {\mn@doi@[]}}
\def\mn@doi@[#1]#2{\def\@tempa{#1}\ifx\@tempa\@empty \href
  {http://dx.doi.org/#2} {doi:#2}\else \href {http://dx.doi.org/#2} {#1}\fi
  \endgroup}
\def\mn@eprint#1#2{\mn@eprint@#1:#2::\@nil}
\def\mn@eprint@arXiv#1{\href {http://arxiv.org/abs/#1} {{\tt arXiv:#1}}}
\def\mn@eprint@dblp#1{\href {http://dblp.uni-trier.de/rec/bibtex/#1.xml}
  {dblp:#1}}
\def\mn@eprint@#1:#2:#3:#4\@nil{\def\@tempa {#1}\def\@tempb {#2}\def\@tempc
  {#3}\ifx \@tempc \@empty \let \@tempc \@tempb \let \@tempb \@tempa \fi \ifx
  \@tempb \@empty \def\@tempb {arXiv}\fi \@ifundefined
  {mn@eprint@\@tempb}{\@tempb:\@tempc}{\expandafter \expandafter \csname
  mn@eprint@\@tempb\endcsname \expandafter{\@tempc}}}

\bibitem[\protect\citeauthoryear{{Aumer}, {White}, {Naab}  \&
  {Scannapieco}}{{Aumer} et~al.}{2013}]{Aumer13}
{Aumer} M.,  {White} S.~D.~M.,  {Naab} T.,   {Scannapieco} C.,  2013, \mn@doi
  [\mnras] {10.1093/mnras/stt1230}, \href
  {http://adsabs.harvard.edu/abs/2013MNRAS.434.3142A} {434, 3142}

\bibitem[\protect\citeauthoryear{{Bah{\'e}} \& {McCarthy}}{{Bah{\'e}} \&
  {McCarthy}}{2015}]{Bahe15}
{Bah{\'e}} Y.~M.,  {McCarthy} I.~G.,  2015, \mn@doi [\mnras]
  {10.1093/mnras/stu2293}, \href
  {http://adsabs.harvard.edu/abs/2015MNRAS.447..969B} {447, 969}

\bibitem[\protect\citeauthoryear{{Behroozi}, {Wechsler}  \& {Wu}}{{Behroozi}
  et~al.}{2013a}]{Behroozi13}
{Behroozi} P.~S.,  {Wechsler} R.~H.,   {Wu} H.-Y.,  2013a, \mn@doi [\apj]
  {10.1088/0004-637X/762/2/109}, \href
  {http://adsabs.harvard.edu/abs/2013ApJ...762..109B} {762, 109}

\bibitem[\protect\citeauthoryear{{Behroozi}, {Wechsler}, {Wu}, {Busha},
  {Klypin}  \& {Primack}}{{Behroozi} et~al.}{2013b}]{consistenttrees}
{Behroozi} P.~S.,  {Wechsler} R.~H.,  {Wu} H.-Y.,  {Busha} M.~T.,  {Klypin}
  A.~A.,   {Primack} J.~R.,  2013b, \mn@doi [\apj]
  {10.1088/0004-637X/763/1/18}, \href
  {http://adsabs.harvard.edu/abs/2013ApJ...763...18B} {763, 18}

\bibitem[\protect\citeauthoryear{{Bezanson}, {van Dokkum}, {Tal}, {Marchesini},
  {Kriek}, {Franx}  \& {Coppi}}{{Bezanson} et~al.}{2009}]{Bezanson09}
{Bezanson} R.,  {van Dokkum} P.~G.,  {Tal} T.,  {Marchesini} D.,  {Kriek} M.,
  {Franx} M.,   {Coppi} P.,  2009, \mn@doi [\apj]
  {10.1088/0004-637X/697/2/1290}, \href
  {http://adsabs.harvard.edu/abs/2009ApJ...697.1290B} {697, 1290}

\bibitem[\protect\citeauthoryear{{Bondi}}{{Bondi}}{1952}]{Bondi52}
{Bondi} H.,  1952, \mn@doi [\mnras] {10.1093/mnras/112.2.195}, \href
  {http://adsabs.harvard.edu/abs/1952MNRAS.112..195B} {112, 195}

\bibitem[\protect\citeauthoryear{{Brennan}, {Choi}, {Somerville}, {Hirschmann},
  {Naab}  \& {Ostriker}}{{Brennan} et~al.}{2018}]{Brennan18}
{Brennan} R.,  {Choi} E.,  {Somerville} R.~S.,  {Hirschmann} M.,  {Naab} T.,
  {Ostriker} J.~P.,  2018, \mn@doi [\apj] {10.3847/1538-4357/aac2c4}, \href
  {http://adsabs.harvard.edu/abs/2018ApJ...860...14B} {860, 14}

\bibitem[\protect\citeauthoryear{{Bruzual} \& {Charlot}}{{Bruzual} \&
  {Charlot}}{2003}]{Bruzual03}
{Bruzual} G.,  {Charlot} S.,  2003, \mn@doi [\mnras]
  {10.1046/j.1365-8711.2003.06897.x}, \href
  {http://adsabs.harvard.edu/abs/2003MNRAS.344.1000B} {344, 1000}

\bibitem[\protect\citeauthoryear{{Buitrago}, {Trujillo}, {Conselice},
  {Bouwens}, {Dickinson}  \& {Yan}}{{Buitrago} et~al.}{2008}]{Buitrago06}
{Buitrago} F.,  {Trujillo} I.,  {Conselice} C.~J.,  {Bouwens} R.~J.,
  {Dickinson} M.,   {Yan} H.,  2008, \mn@doi [\apjl] {10.1086/592836}, \href
  {http://adsabs.harvard.edu/abs/2008ApJ...687L..61B} {687, L61}

\bibitem[\protect\citeauthoryear{{Calderon}, {Berlind}  \& {Sinha}}{{Calderon}
  et~al.}{2018}]{Calderon18}
{Calderon} V.~F.,  {Berlind} A.~A.,   {Sinha} M.,  2018, \mn@doi [\mnras]
  {10.1093/mnras/sty2000}, \href
  {http://adsabs.harvard.edu/abs/2018MNRAS.tmp.1895C} {}

\bibitem[\protect\citeauthoryear{{Cameron}}{{Cameron}}{2011}]{Cameron11}
{Cameron} E.,  2011, \mn@doi [\pasa] {10.1071/AS10046}, \href
  {http://adsabs.harvard.edu/abs/2011PASA...28..128C} {28, 128}

\bibitem[\protect\citeauthoryear{{Campbell}, {van den Bosch}, {Hearin},
  {Padmanabhan}, {Berlind}, {Mo}, {Tinker}  \& {Yang}}{{Campbell}
  et~al.}{2015}]{Campbell15}
{Campbell} D.,  {van den Bosch} F.~C.,  {Hearin} A.,  {Padmanabhan} N.,
  {Berlind} A.,  {Mo} H.~J.,  {Tinker} J.,   {Yang} X.,  2015, \mn@doi [\mnras]
  {10.1093/mnras/stv1091}, \href
  {http://adsabs.harvard.edu/abs/2015MNRAS.452..444C} {452, 444}

\bibitem[\protect\citeauthoryear{{Choi}, {Ostriker}, {Naab}  \&
  {Johansson}}{{Choi} et~al.}{2012}]{Choi12}
{Choi} E.,  {Ostriker} J.~P.,  {Naab} T.,   {Johansson} P.~H.,  2012, \mn@doi
  [\apj] {10.1088/0004-637X/754/2/125}, \href
  {http://adsabs.harvard.edu/abs/2012ApJ...754..125C} {754, 125}

\bibitem[\protect\citeauthoryear{{Choi}, {Naab}, {Ostriker}, {Johansson}  \&
  {Moster}}{{Choi} et~al.}{2014}]{Choi14}
{Choi} E.,  {Naab} T.,  {Ostriker} J.~P.,  {Johansson} P.~H.,   {Moster} B.~P.,
   2014, \mn@doi [\mnras] {10.1093/mnras/stu874}, \href
  {http://adsabs.harvard.edu/abs/2014MNRAS.442..440C} {442, 440}

\bibitem[\protect\citeauthoryear{{Choi}, {Ostriker}, {Naab}, {Oser}  \&
  {Moster}}{{Choi} et~al.}{2015}]{Choi15}
{Choi} E.,  {Ostriker} J.~P.,  {Naab} T.,  {Oser} L.,   {Moster} B.~P.,  2015,
  \mn@doi [\mnras] {10.1093/mnras/stv575}, \href
  {http://adsabs.harvard.edu/abs/2015MNRAS.449.4105C} {449, 4105}

\bibitem[\protect\citeauthoryear{{Choi}, {Ostriker}, {Naab}, {Somerville},
  {Hirschmann}, {N{\'u}{\~n}ez}, {Hu}  \& {Oser}}{{Choi} et~al.}{2017}]{Choi17}
{Choi} E.,  {Ostriker} J.~P.,  {Naab} T.,  {Somerville} R.~S.,  {Hirschmann}
  M.,  {N{\'u}{\~n}ez} A.,  {Hu} C.-Y.,   {Oser} L.,  2017, \mn@doi [\apj]
  {10.3847/1538-4357/aa7849}, \href
  {http://adsabs.harvard.edu/abs/2017ApJ...844...31C} {844, 31}

\bibitem[\protect\citeauthoryear{{Choi}, {Somerville}, {Ostriker}, {Naab}  \&
  {Hirschmann}}{{Choi} et~al.}{2018}]{Choi18}
{Choi} E.,  {Somerville} R.~S.,  {Ostriker} J.~P.,  {Naab} T.,   {Hirschmann}
  M.,  2018, \mn@doi [\apj] {10.3847/1538-4357/aae076}, \href
  {http://adsabs.harvard.edu/abs/2018ApJ...866...91C} {866, 91}

\bibitem[\protect\citeauthoryear{{Cullen} \& {Dehnen}}{{Cullen} \&
  {Dehnen}}{2010}]{Cullen10}
{Cullen} L.,  {Dehnen} W.,  2010, \mn@doi [\mnras]
  {10.1111/j.1365-2966.2010.17158.x}, \href
  {http://adsabs.harvard.edu/abs/2010MNRAS.408..669C} {408, 669}

\bibitem[\protect\citeauthoryear{{Daddi} et~al.,}{{Daddi}
  et~al.}{2005}]{Daddi05}
{Daddi} E.,  et~al., 2005, \mn@doi [\apj] {10.1086/430104}, \href
  {http://adsabs.harvard.edu/abs/2005ApJ...626..680D} {626, 680}

\bibitem[\protect\citeauthoryear{{Dekel}, {Devor}  \& {Hetzroni}}{{Dekel}
  et~al.}{2003}]{Dekel03}
{Dekel} A.,  {Devor} J.,   {Hetzroni} G.,  2003, \mn@doi [\mnras]
  {10.1046/j.1365-8711.2003.06432.x}, \href
  {http://adsabs.harvard.edu/abs/2003MNRAS.341..326D} {341, 326}

\bibitem[\protect\citeauthoryear{{Dubois} et~al.,}{{Dubois}
  et~al.}{2014}]{Dubois14}
{Dubois} Y.,  et~al., 2014, \mn@doi [\mnras] {10.1093/mnras/stu1227}, \href
  {http://adsabs.harvard.edu/abs/2014MNRAS.444.1453D} {444, 1453}

\bibitem[\protect\citeauthoryear{{Durier} \& {Dalla Vecchia}}{{Durier} \&
  {Dalla Vecchia}}{2012}]{Durier12}
{Durier} F.,  {Dalla Vecchia} C.,  2012, \mn@doi [\mnras]
  {10.1111/j.1365-2966.2011.19712.x}, \href
  {http://adsabs.harvard.edu/abs/2012MNRAS.419..465D} {419, 465}

\bibitem[\protect\citeauthoryear{{Farouki} \& {Shapiro}}{{Farouki} \&
  {Shapiro}}{1981}]{Farouki81}
{Farouki} R.,  {Shapiro} S.~L.,  1981, \mn@doi [\apj] {10.1086/158563}, \href
  {http://adsabs.harvard.edu/abs/1981ApJ...243...32F} {243, 32}

\bibitem[\protect\citeauthoryear{{Franx}, {van Dokkum}, {F{\"o}rster
  Schreiber}, {Wuyts}, {Labb{\'e}}  \& {Toft}}{{Franx} et~al.}{2008}]{Franx08}
{Franx} M.,  {van Dokkum} P.~G.,  {F{\"o}rster Schreiber} N.~M.,  {Wuyts} S.,
  {Labb{\'e}} I.,   {Toft} S.,  2008, \mn@doi [\apj] {10.1086/592431}, \href
  {http://adsabs.harvard.edu/abs/2008ApJ...688..770F} {688, 770}

\bibitem[\protect\citeauthoryear{{Frigo}, {Naab}, {Hirschmann}, {Choi},
  {Somerville}, {Krajnovic}, {Dav{\'e}}  \& {Cappellari}}{{Frigo}
  et~al.}{2018}]{Frigo18}
{Frigo} M.,  {Naab} T.,  {Hirschmann} M.,  {Choi} E.,  {Somerville} R.,
  {Krajnovic} D.,  {Dav{\'e}} R.,   {Cappellari} M.,  2018, arXiv e-prints,
  \href {http://adsabs.harvard.edu/abs/2018arXiv181111059F} {}

\bibitem[\protect\citeauthoryear{{Gunn} \& {Gott}}{{Gunn} \&
  {Gott}}{1972}]{Gunn72}
{Gunn} J.~E.,  {Gott} III J.~R.,  1972, \mn@doi [\apj] {10.1086/151605}, \href
  {http://adsabs.harvard.edu/abs/1972ApJ...176....1G} {176, 1}

\bibitem[\protect\citeauthoryear{{Haardt} \& {Madau}}{{Haardt} \&
  {Madau}}{2012}]{Haardt12}
{Haardt} F.,  {Madau} P.,  2012, \mn@doi [\apj] {10.1088/0004-637X/746/2/125},
  \href {http://adsabs.harvard.edu/abs/2012ApJ...746..125H} {746, 125}

\bibitem[\protect\citeauthoryear{{Hearin}, {Watson}  \& {van den
  Bosch}}{{Hearin} et~al.}{2015}]{Hearin15}
{Hearin} A.~P.,  {Watson} D.~F.,   {van den Bosch} F.~C.,  2015, \mn@doi
  [\mnras] {10.1093/mnras/stv1358}, \href
  {http://adsabs.harvard.edu/abs/2015MNRAS.452.1958H} {452, 1958}

\bibitem[\protect\citeauthoryear{{Hearin}, {Behroozi}  \& {van den
  Bosch}}{{Hearin} et~al.}{2016}]{Hearin16}
{Hearin} A.~P.,  {Behroozi} P.~S.,   {van den Bosch} F.~C.,  2016, \mn@doi
  [\mnras] {10.1093/mnras/stw1462}, \href
  {http://adsabs.harvard.edu/abs/2016MNRAS.461.2135H} {461, 2135}

\bibitem[\protect\citeauthoryear{{Hirschmann}, {De Lucia}, {Iovino}  \&
  {Cucciati}}{{Hirschmann} et~al.}{2013}]{Hirschmann13}
{Hirschmann} M.,  {De Lucia} G.,  {Iovino} A.,   {Cucciati} O.,  2013, \mn@doi
  [\mnras] {10.1093/mnras/stt827}, \href
  {http://adsabs.harvard.edu/abs/2013MNRAS.433.1479H} {433, 1479}

\bibitem[\protect\citeauthoryear{{Hirschmann}, {De Lucia}, {Wilman},
  {Weinmann}, {Iovino}, {Cucciati}, {Zibetti}  \& {Villalobos}}{{Hirschmann}
  et~al.}{2014}]{Hirschmann14}
{Hirschmann} M.,  {De Lucia} G.,  {Wilman} D.,  {Weinmann} S.,  {Iovino} A.,
  {Cucciati} O.,  {Zibetti} S.,   {Villalobos} {\'A}.,  2014, \mn@doi [\mnras]
  {10.1093/mnras/stu1609}, \href
  {http://adsabs.harvard.edu/abs/2014MNRAS.444.2938H} {444, 2938}

\bibitem[\protect\citeauthoryear{{Hirschmann}, {Charlot}, {Feltre}, {Naab},
  {Choi}, {Ostriker}  \& {Somerville}}{{Hirschmann}
  et~al.}{2017}]{Hirschmann17}
{Hirschmann} M.,  {Charlot} S.,  {Feltre} A.,  {Naab} T.,  {Choi} E.,
  {Ostriker} J.~P.,   {Somerville} R.~S.,  2017, \mn@doi [\mnras]
  {10.1093/mnras/stx2180}, \href
  {http://adsabs.harvard.edu/abs/2017MNRAS.472.2468H} {472, 2468}

\bibitem[\protect\citeauthoryear{{Hirschmann}, {Charlot}, {Feltre}, {Naab},
  {Somerville}  \& {Choi}}{{Hirschmann} et~al.}{2018}]{Hirschmann18}
{Hirschmann} M.,  {Charlot} S.,  {Feltre} A.,  {Naab} T.,  {Somerville} R.~S.,
   {Choi} E.,  2018, arXiv e-prints, \href
  {http://adsabs.harvard.edu/abs/2018arXiv181107909H} {}

\bibitem[\protect\citeauthoryear{{Hopkins}}{{Hopkins}}{2013}]{Hopkins13}
{Hopkins} P.~F.,  2013, \mn@doi [\mnras] {10.1093/mnras/sts210}, \href
  {http://adsabs.harvard.edu/abs/2013MNRAS.428.2840H} {428, 2840}

\bibitem[\protect\citeauthoryear{{Hu}, {Naab}, {Walch}, {Moster}  \&
  {Oser}}{{Hu} et~al.}{2014}]{Hu14}
{Hu} C.-Y.,  {Naab} T.,  {Walch} S.,  {Moster} B.~P.,   {Oser} L.,  2014,
  \mn@doi [\mnras] {10.1093/mnras/stu1187}, \href
  {http://adsabs.harvard.edu/abs/2014MNRAS.443.1173H} {443, 1173}

\bibitem[\protect\citeauthoryear{{Kauffmann}}{{Kauffmann}}{2015}]{Kauffmann15}
{Kauffmann} G.,  2015, \mn@doi [\mnras] {10.1093/mnras/stv2113}, \href
  {http://adsabs.harvard.edu/abs/2015MNRAS.454.1840K} {454, 1840}

\bibitem[\protect\citeauthoryear{{Kauffmann}}{{Kauffmann}}{2018}]{Kauffmann18}
{Kauffmann} G.,  2018, \mn@doi [\mnras] {10.1093/mnrasl/slx204}, \href
  {http://adsabs.harvard.edu/abs/2018MNRAS.475L..45K} {475, L45}

\bibitem[\protect\citeauthoryear{{Kauffmann}, {Li}  \& {Heckman}}{{Kauffmann}
  et~al.}{2010}]{Kauffmann10}
{Kauffmann} G.,  {Li} C.,   {Heckman} T.~M.,  2010, \mn@doi [\mnras]
  {10.1111/j.1365-2966.2010.17337.x}, \href
  {http://adsabs.harvard.edu/abs/2010MNRAS.409..491K} {409, 491}

\bibitem[\protect\citeauthoryear{{Kauffmann}, {Li}, {Zhang}  \&
  {Weinmann}}{{Kauffmann} et~al.}{2013}]{Kauffmann13}
{Kauffmann} G.,  {Li} C.,  {Zhang} W.,   {Weinmann} S.,  2013, \mn@doi [\mnras]
  {10.1093/mnras/stt007}, \href
  {http://adsabs.harvard.edu/abs/2013MNRAS.430.1447K} {430, 1447}

\bibitem[\protect\citeauthoryear{{Kimm} et~al.,}{{Kimm} et~al.}{2009}]{Kimm09}
{Kimm} T.,  et~al., 2009, \mn@doi [\mnras] {10.1111/j.1365-2966.2009.14414.x},
  \href {http://adsabs.harvard.edu/abs/2009MNRAS.394.1131K} {394, 1131}

\bibitem[\protect\citeauthoryear{{Knobel}, {Lilly}, {Woo}  \& {Kova{\v
  c}}}{{Knobel} et~al.}{2015}]{Knobel15}
{Knobel} C.,  {Lilly} S.~J.,  {Woo} J.,   {Kova{\v c}} K.,  2015, \mn@doi
  [\apj] {10.1088/0004-637X/800/1/24}, \href
  {http://adsabs.harvard.edu/abs/2015ApJ...800...24K} {800, 24}

\bibitem[\protect\citeauthoryear{{Larson}, {Tinsley}  \& {Caldwell}}{{Larson}
  et~al.}{1980}]{Larson80}
{Larson} R.~B.,  {Tinsley} B.~M.,   {Caldwell} C.~N.,  1980, \mn@doi [\apj]
  {10.1086/157917}, \href {http://adsabs.harvard.edu/abs/1980ApJ...237..692L}
  {237, 692}

\bibitem[\protect\citeauthoryear{{Lubin}, {Gal}, {Lemaux}, {Kocevski}  \&
  {Squires}}{{Lubin} et~al.}{2009}]{Lubin2009}
{Lubin} L.~M.,  {Gal} R.~R.,  {Lemaux} B.~C.,  {Kocevski} D.~D.,   {Squires}
  G.~K.,  2009, \mn@doi [\aj] {10.1088/0004-6256/137/6/4867}, \href
  {http://adsabs.harvard.edu/abs/2009AJ....137.4867L} {137, 4867}

\bibitem[\protect\citeauthoryear{{Moore}, {Lake}  \& {Katz}}{{Moore}
  et~al.}{1998}]{Moore98}
{Moore} B.,  {Lake} G.,   {Katz} N.,  1998, \mn@doi [\apj] {10.1086/305264},
  \href {http://adsabs.harvard.edu/abs/1998ApJ...495..139M} {495, 139}

\bibitem[\protect\citeauthoryear{{Naab} \& {Ostriker}}{{Naab} \&
  {Ostriker}}{2017}]{NaabOstriker17}
{Naab} T.,  {Ostriker} J.~P.,  2017, \mn@doi [\araa]
  {10.1146/annurev-astro-081913-040019}, \href
  {http://adsabs.harvard.edu/abs/2017ARA%26A..55...59N} {55, 59}

\bibitem[\protect\citeauthoryear{{Naab}, {Johansson}, {Ostriker}  \&
  {Efstathiou}}{{Naab} et~al.}{2007}]{Naab07}
{Naab} T.,  {Johansson} P.~H.,  {Ostriker} J.~P.,   {Efstathiou} G.,  2007,
  \mn@doi [\apj] {10.1086/510841}, \href
  {http://adsabs.harvard.edu/abs/2007ApJ...658..710N} {658, 710}

\bibitem[\protect\citeauthoryear{{N{\'u}{\~n}ez}, {Ostriker}, {Naab}, {Oser},
  {Hu}  \& {Choi}}{{N{\'u}{\~n}ez} et~al.}{2017}]{Nunez17}
{N{\'u}{\~n}ez} A.,  {Ostriker} J.~P.,  {Naab} T.,  {Oser} L.,  {Hu} C.-Y.,
  {Choi} E.,  2017, \mn@doi [\apj] {10.3847/1538-4357/836/2/204}, \href
  {http://adsabs.harvard.edu/abs/2017ApJ...836..204N} {836, 204}

\bibitem[\protect\citeauthoryear{{Nyman} et~al.,}{{Nyman}
  et~al.}{1992}]{Nyman92}
{Nyman} L.-A.,  et~al., 1992, \aaps, \href
  {http://adsabs.harvard.edu/abs/1992A%26AS...93..121N} {93, 121}

\bibitem[\protect\citeauthoryear{{Oser}, {Ostriker}, {Naab}, {Johansson}  \&
  {Burkert}}{{Oser} et~al.}{2010}]{Oser10}
{Oser} L.,  {Ostriker} J.~P.,  {Naab} T.,  {Johansson} P.~H.,   {Burkert} A.,
  2010, \mn@doi [\apj] {10.1088/0004-637X/725/2/2312}, \href
  {http://adsabs.harvard.edu/abs/2010ApJ...725.2312O} {725, 2312}

\bibitem[\protect\citeauthoryear{{Oser}, {Naab}, {Ostriker}  \&
  {Johansson}}{{Oser} et~al.}{2012}]{Oser12}
{Oser} L.,  {Naab} T.,  {Ostriker} J.~P.,   {Johansson} P.~H.,  2012, \mn@doi
  [\apj] {10.1088/0004-637X/744/1/63}, \href
  {http://adsabs.harvard.edu/abs/2012ApJ...744...63O} {744, 63}

\bibitem[\protect\citeauthoryear{{Ostriker}, {Choi}, {Ciotti}, {Novak}  \&
  {Proga}}{{Ostriker} et~al.}{2010}]{Ostriker2010}
{Ostriker} J.~P.,  {Choi} E.,  {Ciotti} L.,  {Novak} G.~S.,   {Proga} D.,
  2010, \mn@doi [\apj] {10.1088/0004-637X/722/1/642}, \href
  {https://ui.adsabs.harvard.edu/abs/2010ApJ...722..642O} {722, 642}

\bibitem[\protect\citeauthoryear{{Peng} et~al.,}{{Peng} et~al.}{2010}]{Peng10}
{Peng} Y.-j.,  et~al., 2010, \mn@doi [\apj] {10.1088/0004-637X/721/1/193},
  \href {http://adsabs.harvard.edu/abs/2010ApJ...721..193P} {721, 193}

\bibitem[\protect\citeauthoryear{{Peng}, {Lilly}, {Renzini}  \&
  {Carollo}}{{Peng} et~al.}{2012}]{Peng12}
{Peng} Y.-j.,  {Lilly} S.~J.,  {Renzini} A.,   {Carollo} M.,  2012, \mn@doi
  [\apj] {10.1088/0004-637X/757/1/4}, \href
  {http://adsabs.harvard.edu/abs/2012ApJ...757....4P} {757, 4}

\bibitem[\protect\citeauthoryear{{Phillips}, {Wheeler}, {Boylan-Kolchin},
  {Bullock}, {Cooper}  \& {Tollerud}}{{Phillips} et~al.}{2014}]{Phillips14}
{Phillips} J.~I.,  {Wheeler} C.,  {Boylan-Kolchin} M.,  {Bullock} J.~S.,
  {Cooper} M.~C.,   {Tollerud} E.~J.,  2014, \mn@doi [\mnras]
  {10.1093/mnras/stt2023}, \href
  {http://adsabs.harvard.edu/abs/2014MNRAS.437.1930P} {437, 1930}

\bibitem[\protect\citeauthoryear{{Porter}, {Somerville}, {Primack}, {Croton},
  {Covington}, {Graves}  \& {Faber}}{{Porter} et~al.}{2014}]{Porter14}
{Porter} L.~A.,  {Somerville} R.~S.,  {Primack} J.~R.,  {Croton} D.~J.,
  {Covington} M.~D.,  {Graves} G.~J.,   {Faber} S.~M.,  2014, \mn@doi [\mnras]
  {10.1093/mnras/stu1701}, \href
  {http://adsabs.harvard.edu/abs/2014MNRAS.445.3092P} {445, 3092}

\bibitem[\protect\citeauthoryear{{Read} \& {Hayfield}}{{Read} \&
  {Hayfield}}{2012}]{Read12}
{Read} J.~I.,  {Hayfield} T.,  2012, \mn@doi [\mnras]
  {10.1111/j.1365-2966.2012.20819.x}, \href
  {http://adsabs.harvard.edu/abs/2012MNRAS.422.3037R} {422, 3037}

\bibitem[\protect\citeauthoryear{{R{\"o}ttgers} \& {Arth}}{{R{\"o}ttgers} \&
  {Arth}}{2018}]{pygad}
{R{\"o}ttgers} B.,  {Arth} A.,  2018, preprint, \href
  {http://adsabs.harvard.edu/abs/2018arXiv180303652R} {} (\mn@eprint {arXiv}
  {1803.03652})

\bibitem[\protect\citeauthoryear{{Saitoh} \& {Makino}}{{Saitoh} \&
  {Makino}}{2009}]{Saitoh09}
{Saitoh} T.~R.,  {Makino} J.,  2009, \mn@doi [\apjl]
  {10.1088/0004-637X/697/2/L99}, \href
  {http://adsabs.harvard.edu/abs/2009ApJ...697L..99S} {697, L99}

\bibitem[\protect\citeauthoryear{{Sales}, {Navarro}, {Theuns}, {Schaye},
  {White}, {Frenk}, {Crain}  \& {Dalla Vecchia}}{{Sales}
  et~al.}{2012}]{Sales12}
{Sales} L.~V.,  {Navarro} J.~F.,  {Theuns} T.,  {Schaye} J.,  {White} S.~D.~M.,
   {Frenk} C.~S.,  {Crain} R.~A.,   {Dalla Vecchia} C.,  2012, \mn@doi [\mnras]
  {10.1111/j.1365-2966.2012.20975.x}, \href
  {http://adsabs.harvard.edu/abs/2012MNRAS.423.1544S} {423, 1544}

\bibitem[\protect\citeauthoryear{{Sazonov}, {Ostriker}  \& {Sunyaev}}{{Sazonov}
  et~al.}{2004}]{Sazonov04}
{Sazonov} S.~Y.,  {Ostriker} J.~P.,   {Sunyaev} R.~A.,  2004, \mn@doi [\mnras]
  {10.1111/j.1365-2966.2004.07184.x}, \href
  {http://adsabs.harvard.edu/abs/2004MNRAS.347..144S} {347, 144}

\bibitem[\protect\citeauthoryear{{Sazonov}, {Ostriker}, {Ciotti}  \&
  {Sunyaev}}{{Sazonov} et~al.}{2005}]{Sazonov05}
{Sazonov} S.~Y.,  {Ostriker} J.~P.,  {Ciotti} L.,   {Sunyaev} R.~A.,  2005,
  \mn@doi [\mnras] {10.1111/j.1365-2966.2005.08763.x}, \href
  {http://adsabs.harvard.edu/abs/2005MNRAS.358..168S} {358, 168}

\bibitem[\protect\citeauthoryear{{Shen} et~al.,}{{Shen}
  et~al.}{2019}]{Shen2019}
{Shen} L.,  et~al., 2019, \mn@doi [\mnras] {10.1093/mnras/stz152}, \href
  {http://adsabs.harvard.edu/abs/2019MNRAS.484.2433S} {484, 2433}

\bibitem[\protect\citeauthoryear{{Sin}, {Lilly}  \& {Henriques}}{{Sin}
  et~al.}{2017}]{Sin17}
{Sin} L.~P.~T.,  {Lilly} S.~J.,   {Henriques} B.~M.~B.,  2017, \mn@doi [\mnras]
  {10.1093/mnras/stx1674}, \href
  {http://adsabs.harvard.edu/abs/2017MNRAS.471.1192S} {471, 1192}

\bibitem[\protect\citeauthoryear{{Somerville} \& {Dav{\'e}}}{{Somerville} \&
  {Dav{\'e}}}{2015}]{Somerville15}
{Somerville} R.~S.,  {Dav{\'e}} R.,  2015, \mn@doi [\araa]
  {10.1146/annurev-astro-082812-140951}, \href
  {http://adsabs.harvard.edu/abs/2015ARA%26A..53...51S} {53, 51}

\bibitem[\protect\citeauthoryear{{Spergel} et~al.,}{{Spergel}
  et~al.}{2007}]{Spergel}
{Spergel} D.~N.,  et~al., 2007, \mn@doi [\apjs] {10.1086/513700}, \href
  {http://adsabs.harvard.edu/abs/2007ApJS..170..377S} {170, 377}

\bibitem[\protect\citeauthoryear{{Springel}}{{Springel}}{2005}]{Springel05}
{Springel} V.,  2005, \mn@doi [\mnras] {10.1111/j.1365-2966.2005.09655.x},
  \href {http://adsabs.harvard.edu/abs/2005MNRAS.364.1105S} {364, 1105}

\bibitem[\protect\citeauthoryear{{Szomoru}, {Franx}  \& {van Dokkum}}{{Szomoru}
  et~al.}{2012}]{Szomoru09}
{Szomoru} D.,  {Franx} M.,   {van Dokkum} P.~G.,  2012, \mn@doi [\apj]
  {10.1088/0004-637X/749/2/121}, \href
  {http://adsabs.harvard.edu/abs/2012ApJ...749..121S} {749, 121}

\bibitem[\protect\citeauthoryear{{Tinker}, {Hahn}, {Mao}, {Wetzel}  \&
  {Conroy}}{{Tinker} et~al.}{2018}]{Tinker18}
{Tinker} J.~L.,  {Hahn} C.,  {Mao} Y.-Y.,  {Wetzel} A.~R.,   {Conroy} C.,
  2018, \mn@doi [\mnras] {10.1093/mnras/sty666}, \href
  {http://adsabs.harvard.edu/abs/2018MNRAS.477..935T} {477, 935}

\bibitem[\protect\citeauthoryear{{Tonnesen} \& {Bryan}}{{Tonnesen} \&
  {Bryan}}{2009}]{Tonessen09}
{Tonnesen} S.,  {Bryan} G.~L.,  2009, \mn@doi [\apj]
  {10.1088/0004-637X/694/2/789}, \href
  {http://adsabs.harvard.edu/abs/2009ApJ...694..789T} {694, 789}

\bibitem[\protect\citeauthoryear{{Trujillo} et~al.,}{{Trujillo}
  et~al.}{2006}]{Trujillo06}
{Trujillo} I.,  et~al., 2006, \mn@doi [\apj] {10.1086/506464}, \href
  {http://adsabs.harvard.edu/abs/2006ApJ...650...18T} {650, 18}

\bibitem[\protect\citeauthoryear{{Uchiyama} et~al.,}{{Uchiyama}
  et~al.}{2019}]{Uchiyama2019}
{Uchiyama} H.,  et~al., 2019, \mn@doi [\apj] {10.3847/1538-4357/aaef7b}, \href
  {http://adsabs.harvard.edu/abs/2019ApJ...870...45U} {870, 45}

\bibitem[\protect\citeauthoryear{{Vogelsberger} et~al.,}{{Vogelsberger}
  et~al.}{2014}]{Illustris}
{Vogelsberger} M.,  et~al., 2014, \mn@doi [\mnras] {10.1093/mnras/stu1536},
  \href {http://adsabs.harvard.edu/abs/2014MNRAS.444.1518V} {444, 1518}

\bibitem[\protect\citeauthoryear{{Wang} \& {White}}{{Wang} \&
  {White}}{2012}]{Wang12}
{Wang} W.,  {White} S.~D.~M.,  2012, \mn@doi [\mnras]
  {10.1111/j.1365-2966.2012.21256.x}, \href
  {http://adsabs.harvard.edu/abs/2012MNRAS.424.2574W} {424, 2574}

\bibitem[\protect\citeauthoryear{{Wechsler} \& {Tinker}}{{Wechsler} \&
  {Tinker}}{2018}]{Wechsler18}
{Wechsler} R.~H.,  {Tinker} J.~L.,  2018, \mn@doi [\araa]
  {10.1146/annurev-astro-081817-051756}, \href
  {http://adsabs.harvard.edu/abs/2018ARA%26A..56..435W} {56, 435}

\bibitem[\protect\citeauthoryear{{Weinmann}, {van den Bosch}, {Yang}  \&
  {Mo}}{{Weinmann} et~al.}{2006}]{Weinmann06}
{Weinmann} S.~M.,  {van den Bosch} F.~C.,  {Yang} X.,   {Mo} H.~J.,  2006,
  \mn@doi [\mnras] {10.1111/j.1365-2966.2005.09865.x}, \href
  {http://adsabs.harvard.edu/abs/2006MNRAS.366....2W} {366, 2}

\bibitem[\protect\citeauthoryear{{Wetzel} \& {White}}{{Wetzel} \&
  {White}}{2010}]{Wetzel10}
{Wetzel} A.~R.,  {White} M.,  2010, \mn@doi [\mnras]
  {10.1111/j.1365-2966.2009.16191.x}, \href
  {http://adsabs.harvard.edu/abs/2010MNRAS.403.1072W} {403, 1072}

\bibitem[\protect\citeauthoryear{{Wiersma}, {Schaye}  \& {Smith}}{{Wiersma}
  et~al.}{2009}]{Wiersma09}
{Wiersma} R.~P.~C.,  {Schaye} J.,   {Smith} B.~D.,  2009, \mn@doi [\mnras]
  {10.1111/j.1365-2966.2008.14191.x}, \href
  {http://adsabs.harvard.edu/abs/2009MNRAS.393...99W} {393, 99}

\bibitem[\protect\citeauthoryear{{Woo} et~al.,}{{Woo} et~al.}{2013}]{Woo2013}
{Woo} J.,  et~al., 2013, \mn@doi [\mnras] {10.1093/mnras/sts274}, \href
  {http://adsabs.harvard.edu/abs/2013MNRAS.428.3306W} {428, 3306}

\bibitem[\protect\citeauthoryear{{van Dokkum} et~al.,}{{van Dokkum}
  et~al.}{2010}]{vanDokkum10}
{van Dokkum} P.~G.,  et~al., 2010, \mn@doi [\apj]
  {10.1088/0004-637X/709/2/1018}, \href
  {http://adsabs.harvard.edu/abs/2010ApJ...709.1018V} {709, 1018}

\makeatother
\end{thebibliography}


\appendix
\section{Resolution study}
\label{appendix}
We test the convergence with respect to resolution using the wAGN and noAGN simulations of two halos that have been run with eight times better mass resolution than the reference resolution, with $m_{*,\rm{gas}}=5.3\, \times 10^5 \, \rm{h}^{-1}{\rm M}_{\odot}$ and $m_{\rm{dm}}=3.1 \times 10^6 \, \rm{h}^{-1}{\rm M}_{\odot}$, and twice better spatial resolution with $\epsilon_{\rm gas,star} = 200 \, \rm{pc}\, \rm{h}^{-1}$ and  $\epsilon_{\rm halo} = 450 \,\rm{pc}\, \rm{h}^{-1}$. One sees that the difference in quiescent and gas poor fractions holds for the higher resolution simulation.

\begin{figure}
    \centering
 \includegraphics[width=\columnwidth]{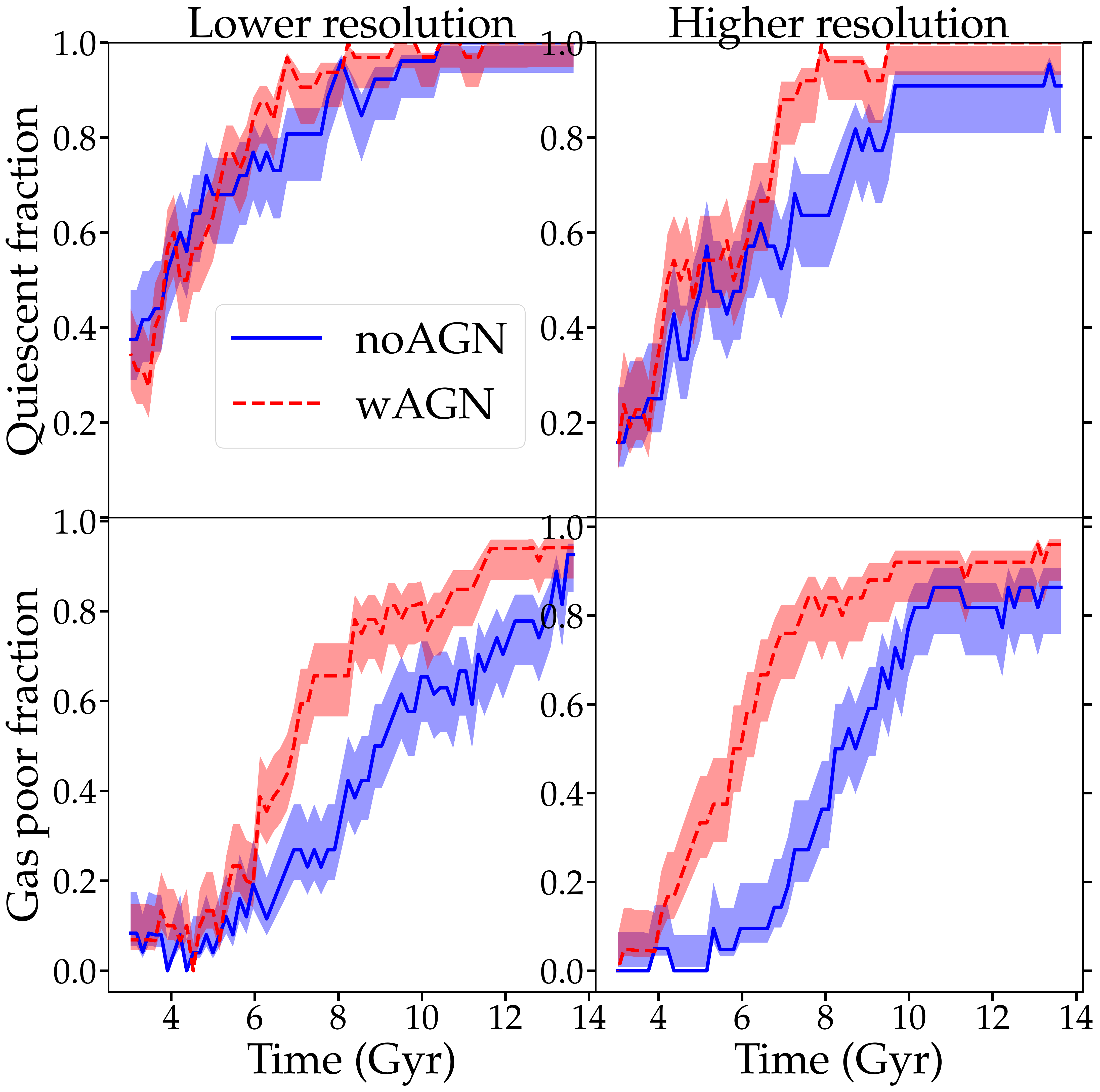}
    \caption{Test of the convergence with respect to resolution using the wAGN and noAGN simulations of two halos that have been run with eight times better mass resolution than the reference resolution, with $m_{*,\rm{gas}}=5.3\, \times 10^5 \, \rm{h}^{-1}{\rm M}_{\odot}$ and $m_{\rm{dm}}=3.1 \times 10^6 \, \rm{h}^{-1}{\rm M}_{\odot}$, and twice better spatial resolution with $\epsilon_{\rm gas,star} = 200 \, \rm{pc}\, \rm{h}^{-1}$ and  $\epsilon_{\rm halo} = 450 \,\rm{pc}\, \rm{h}^{-1}$. The upper panels show the quiescent fractions of satellites galaxies as a function of time, for the lower (left) and the higher resolution simulation (right). The lower panels show the gas poor fractions for the lower (left) and the higher resolution simulation (right)}
    \label{convergence}
\end{figure}


\bsp	
\label{lastpage}
\end{document}